\documentclass[fleqn,usenatbib]{mnras}

\usepackage{newtxtext,newtxmath}
\usepackage[T1]{fontenc}
\usepackage{ae,aecompl}

\usepackage{graphicx}	% Including figure files
\usepackage{amsmath}	% Advanced maths commands
\usepackage{amssymb}	% Extra maths symbols
\usepackage{color}	    % For commenting (TCS)
\usepackage{pdflscape}	% For landscape tables (TCS)
\usepackage{longtable}  % For a table over several pages (FSc)
\usepackage[dvipsnames]{xcolor}
\usepackage{tablefootnote}
\usepackage{ulem}
\usepackage{comment}
\usepackage{dcolumn}

\definecolor{orange}{rgb}{1,0.5,0}

\newcommand{\cco}{$^{13}$CO\,(2\,--\,1)}
\newcommand{\coo}{C$^{18}$O\,(2\,--\,1)}
\newcommand{\kms}{km\,s$^{-1}$}
\newcommand{\vlsr}{v$_{\rm lsr}$}
\newcommand{\tmb}{$T_{\rm mb}$}

\newcommand{\adcprep}{Duarte-Cabral et al.\ submitted}

\newcommand{\sag}{Sagittarius}
\newcommand{\nor}{Norma}
\newcommand{\per}{Perseus}
\newcommand{\scu}{Scutum-Centaurus}

\newcommand{\hii}{\ion{H}{ii}}

\newcolumntype{d}[1]{D{.}{\cdot}{#1}}

\newcolumntype{.}{D{.}{.}{-1}}

\title[SEDIGISM first data release]{The SEDIGISM survey: first data release and overview of the Galactic structure\thanks{This publication is based on data acquired with the Atacama Pathfinder Experiment (APEX), projects 092.F-9315 and 193.C-0584. APEX is a collaboration between the Max-Planck-Institut f\"ur Radioastronomie, the European Southern Observatory, and the Onsala Space Observatory.}}

\author[F. Schuller et al.]{F.\ Schuller$^{1,2,3}$\thanks{E-mail: schuller@mpifr.de},
J.\,S.\,Urquhart$^{4}$, T.\,Csengeri$^{5,1}$,
D.\,Colombo$^{1}$, A.\,Duarte-Cabral$^{6}$,
\newauthor
M.\,Mattern$^{1}$, A.\,Ginsburg$^{7}$,
A.\,R.\,Pettitt$^{8}$, F.\,Wyrowski$^{1}$,
L.\,Anderson$^{9}$, F.\,Azagra$^{2}$,
\newauthor
P.\,Barnes$^{10}$, M.\,Beltran$^{11}$,
H.\,Beuther$^{12}$, S.\,Billington$^{4}$,
L.\, Bronfman$^{13}$, R.\,Cesaroni$^{11}$,
\newauthor
C.\,Dobbs$^{14}$, D.\,Eden$^{15}$, M.-Y.\,Lee$^{16}$, 
S.-N.\,X.\,Medina$^{1}$, K.\,M.\,Menten$^{1}$, T.\,Moore$^{15}$,
\newauthor
F.\,M.\,Montenegro-Montes$^{2}$, 
S.\,Ragan$^{6}$, A.\,Rigby$^{6}$,
M.\,Riener$^{12}$, D. Russeil$^{17}$, 
\newauthor
E.\,Schisano$^{19}$, A.\,Sanchez-Monge$^{18}$, A.\,Traficante$^{19}$, A.\,Zavagno$^{17}$, C.\,Agurto$^{2}$, 
\newauthor
S.\,Bontemps$^{5}$, R.\,Finger$^{13}$, A.\,Giannetti$^{20}$,
E.\,Gonzalez$^{2}$, A.\,K.\,Hernandez$^{21}$, 
\newauthor
T.\,Henning$^{12}$, J.\,Kainulainen$^{22}$, J.\,Kauffmann$^{23}$,
S.\,Leurini$^{24}$, S.\,Lopez$^{7}$, F.\,Mac-Auliffe$^{2}$, 
\newauthor
P.\,Mazumdar$^{1}$, S.\,Molinari$^{19}$,
F.\,Motte$^{25}$, E.\,Muller$^{26}$, Q.\,Nguyen-Luong$^{27}$, 
\newauthor
R.\,Parra$^{2}$,
J.-P.\,Perez-Beaupuits$^{2}$, P.\,Schilke$^{18}$, N.\,Schneider$^{18}$, 
S.\,Suri$^{18}$, L.\,Testi$^{28}$,
\newauthor
K.\,Torstensson$^{2}$,  V.\,S.\,Veena$^{18}$, 
P.\,Venegas$^{2}$, K.\,Wang$^{29}$, M.\,Wienen$^{14,1}$ \\
\\
Affiliations can be found after the references.}

\date{Accepted XXX. Received YYY; in original form ZZZ}

\pubyear{2020}

\begin{document}
\label{firstpage}
\pagerange{\pageref{firstpage}--\pageref{lastpage}}
\maketitle

\begin{abstract}
The SEDIGISM (Structure, Excitation and Dynamics of the Inner Galactic Interstellar Medium) survey used the APEX telescope to map 84~deg$^2$ of the Galactic plane between $\ell = -60$\degr\ and $\ell = +31$\degr\ in several molecular transitions, including \cco\ and \coo, thus probing the moderately dense ($\sim$10$^3$~cm$^{-3}$) component of the interstellar medium. With an angular resolution of 30$''$ and a typical $1\sigma$ sensitivity of 0.8--1.0\,K at 0.25\,\kms\ velocity resolution, it gives access to a wide range of structures, from individual star-forming clumps to giant molecular clouds and complexes.
The coverage includes a good fraction of the first and fourth Galactic quadrants, allowing us to constrain the large scale distribution of cold molecular gas in the inner Galaxy.
In this paper we provide an updated overview of the full survey and the data reduction procedures used. We also assess the quality of these data and describe the data products that are being made publicly available as part of this first data release (DR1).
We present integrated maps and position-velocity maps of the molecular gas and use these to investigate the correlation between the molecular gas and the large scale structural features of the Milky Way such as the spiral arms, Galactic bar and Galactic centre.
We find that approximately 60 per cent of the molecular gas is associated with the spiral arms and these appear as strong intensity peaks in the derived Galactocentric distribution. We also find strong peaks in intensity at specific longitudes that correspond to the Galactic centre and well known star forming complexes, revealing that the $^{13}$CO emission is concentrated in a small number of complexes rather than evenly distributed along spiral arms.
\end{abstract}


%%%%%%%%%%%%%%%%%%%%%%%%%%%%%%%%%%%%%%%%%%%%%%%%%%
%%%%%%%%%%%%%%%%% BODY OF PAPER %%%%%%%%%%%%%%%%%%

\section{Introduction}

\begin{table*}
\begin{minipage}{\linewidth}
\begin{center}
  \caption{Main characteristics of recent, large scale CO surveys of the inner Galactic plane.}\label{tab:surveys}
    % \centering
    \begin{tabular}{lllllll}
    \hline
    Survey  &  Transitions  & Coverage  & Angular  & Velocity  & rms  & Reference  \\
    Name  & & & resolution  & resolution  & (\tmb)  & \\
    \hline
    GRS      & $^{13}$CO (J=1--0) & 18\degr $\leq$ $\ell$ $\leq$ 55.7\degr, & 46$''$ & 0.21~\kms & $\sim$0.2~K  & \citet{jackson2006} \\ 
     & & $\vert b \vert$ $\leq$ 1.0\degr\ & & & & \\
    CHaMP  &  CO/$^{13}$CO/C$^{18}$O/  & 280\degr $\leq$ $\ell$ $\leq$ 300\degr, & 37--40$''$ & 0.08-0.10~\kms & 0.4-0.7 K & \citet{Barnes2011} \\
     & CN/N$_2$H$^+$/HNC/ & -4\degr\ $\leq$ b $\leq$ +2\degr\ & & & & \\
     & HCO$^+$/HCN (J=1--0) & & & & & \\
    COHRS    & $^{12}$CO (J=3--2)  & 10\degr $\leq$ $\ell$ $\leq$ 65\degr, & 16$''$  & 1~\kms  & $\sim$1~K  & \citet{dempsey2013} \\ 
     & & $\vert b \vert$ $\leq$ 0.5\degr\ & & & & \\
    CHIMPS   & $^{13}$CO/C$^{18}$O   & 28\degr $\leq$ $\ell$ $\leq$ 46\degr, & 15$''$  & 0.5~\kms  & $\sim$0.6~K & \citet{rigby2016} \\ 
     & (J=3--2) & $\vert b \vert$ $\leq$ 0.5\degr\ & & & & \\
    FUGIN    & CO/$^{13}$CO/C$^{18}$O  & 10\degr $\leq$ $\ell$ $\leq$ 50\degr, $\vert b \vert$ $\leq$ 1.0\degr\ & 20$''$ & 1.3~\kms & 0.3--0.5~K & \citet{FUGIN2017} \\ 
     & (J=1--0) & 198\degr $\leq$ $\ell$ $\leq$ 236\degr, $\vert b \vert$ $\leq$ 1.0\degr\ & & & & \\
    Mopra-CO & CO/$^{13}$CO/C$^{18}$O/  & 305\degr $\leq$ $\ell$ $\leq$ 345\degr, & 35$''$ & 0.1~\kms & 0.7--1.5~K & \citet{burton2013} \\ 
     & C$^{17}$O (J=1--0) & $\vert b \vert$ $\leq$ 0.5\degr\ & & & & \\
    ThrUMMS  & CO/$^{13}$CO/C$^{18}$O/  & 300\degr $\leq$ $\ell$ $\leq$ 360\degr, & 72$''$ & 0.3~\kms & 0.7--1.3~K & \citet{ref-thrumms} \\ 
     & CN (J=1--0) & $\vert b \vert$ $\leq$ 1.0\degr\ & & & & \\
    MWISP\footnotemark   & CO/$^{13}$CO/C$^{18}$O  & -10\degr $\leq$ $\ell$ $\leq$ 250\degr, & 50$''$ & 0.16~\kms & 0.3--0.5~K & \citet{ref-mwisp} \\ 
     & (J=1--0) & $\vert b \vert$ $\leq$ 5.2\degr\ & & & & \\
    FQS  &  $^{12}$CO/$^{13}$CO   & 220 \degr $\leq$ $\ell$ $\leq$ 240\degr, & 55$''$  & 0.26/0.65~\kms  &  0.3--1.3~K  &  \citet{ref-FQS} \\
     & (J=1--0) & -2.5\degr\ $\leq$ b $\leq$ 0\degr\ & & & & \\
    \hline
    SEDIGISM & $^{13}$CO/C$^{18}$O   & $-$60\degr $\leq$ $\ell$ $\leq$ +18\degr, & 30$''$ & 0.25~\kms & 0.8--1.1~K  & \citet{Schuller2017} \\ 
     & (J=2--1) & $\vert b \vert$ $\leq$ 0.5\degr\ + W43 & & & & \\
    \hline
    \end{tabular}
\end{center}
\end{minipage}
\footnotesize{$^1$ Observations for the MWISP survey are still ongoing; the \citet{ref-mwisp} paper focuses on a region covering 25.8\degr $\leq$ $\ell$ $\leq$ +49.7\degr. }
\end{table*}

\begin{figure*}
    \includegraphics[width=0.90\textwidth]{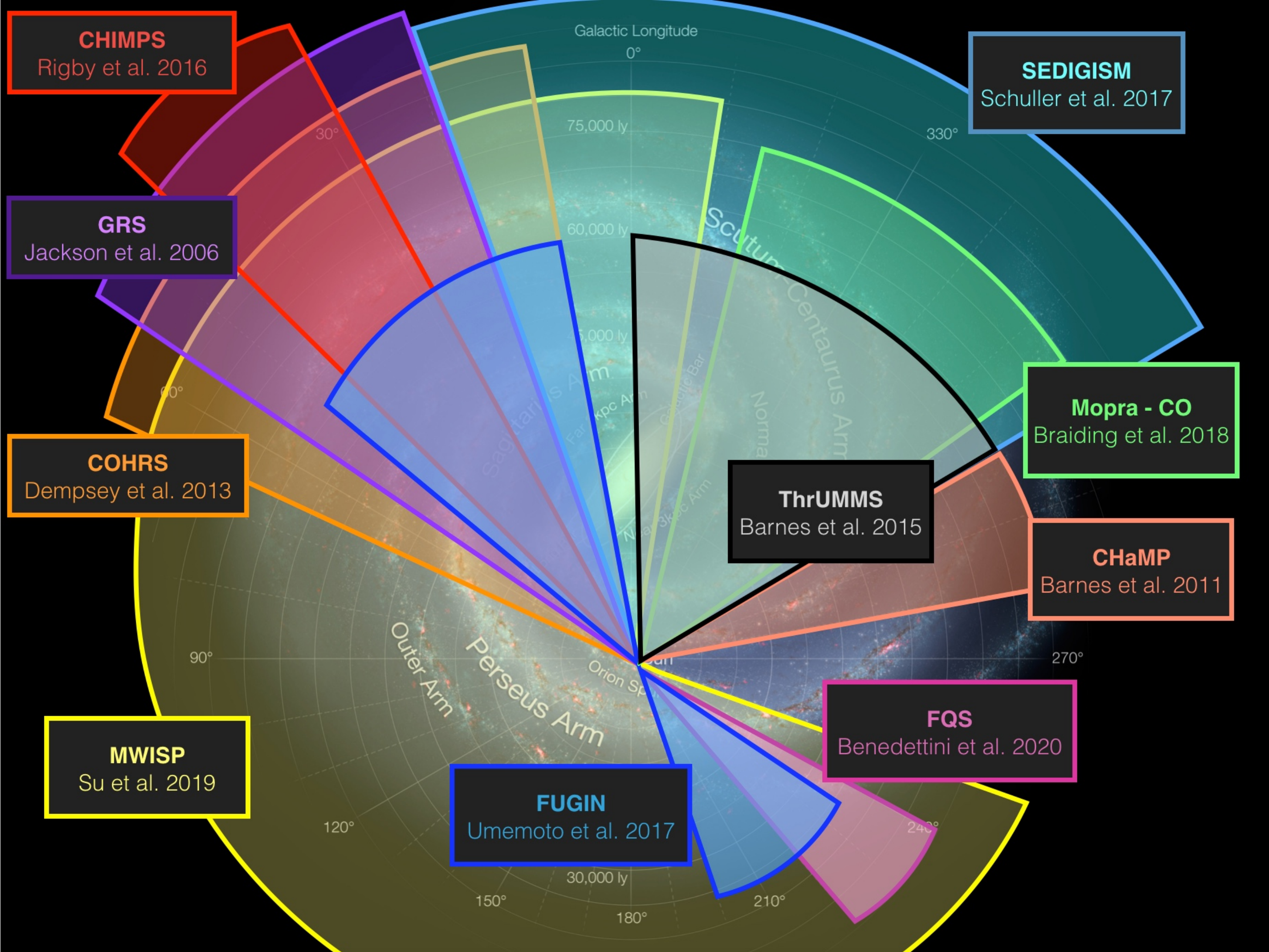}
   \caption{Milky Way coverage of recent CO surveys described in Table\,\ref{tab:surveys}. The background image is a schematic of the Galactic disc as viewed from the Northern Galactic Pole (courtesy of NASA/JPL-Caltech/R. Hurt (SSC/Caltech)) showing the known large-scale features of the Milky Way, such as the spiral arms and the Galactic bar. The survey wedges emanate from the position of the Sun; their respective lengths are arbitrary and do not reflect the sensitivity of each survey.}
    \label{fig:mw_surveys}
\end{figure*}

%%%%%%%%%%%%%%%%%%%%%% Main text starts here
% Context...
Over the last few decades, many systematic continuum surveys of the Galactic plane have been carried out over the full electromagnetic spectrum, from the infrared (e.g.\ GLIMPSE, \citealp{churchwell2009}, MIPSGAL, \citealp{carey2009}, Hi-GAL, \citealp{Molinari2010}) to the (sub-)millimetre (ATLASGAL, \citealp{schuller2009}, BGPS, \citealp{ref-bgps}, JPS, \citealp{moore2015,Eden2017}), and radio range (CORNISH, \citealp{hoare2012,purcell2013}, THOR, \citealp{thor2016,Wang2020}, GLOSTAR, \citealp{medina2019}).
These continuum surveys have been complemented by spectral line surveys, which are essential for estimating  distances and determining physical properties. Some examples include the $^{12}$CO\,(1-0) survey from \citet{dame2001},
the CO Boston University-Five College Radio Astronomy Observatory Galactic Ring Survey \citep[GRS,][]{jackson2006}, the Census of High- and Medium-mass Protostars \citep[CHaMP,][]{Barnes2011}, the CO High-Resolution Survey \citep[COHRS,][]{dempsey2013}, the CO Heterodyne Inner Milky Way Plane Survey \citep[CHIMPS,][]{rigby2016, rigby2019}, the FOREST unbiased Galactic plane imaging survey with Nobeyama \citep[FUGIN,][]{FUGIN2017}, the Mopra Southern Galactic Plane CO Survey \citep{burton2013,MopraCO2018}, the Three-mm Ultimate Mopra Milky Way Survey \citep[ThrUMMS,][]{ref-thrumms}, the Milky Way Imaging Scroll Painting \citep[MWISP,][]{ref-mwisp}, and the Forgotten Quandrant Survey \citep{ref-FQS}.
In Table\,\ref{tab:surveys} and Fig.\,\ref{fig:mw_surveys}, we present a summary of these recently completed CO surveys and show their coverage on a schematic top-down view of the Milky Way.

This wealth of data has greatly enhanced our view of the Galactic structure and its major components. However, there is still no consensus on the exact structure of the Galaxy. For instance, it is not firmly established how many spiral arms are present and what is their exact location \citep[e.g.][]{taylor1993,reid2014,reid2016,Vallee2017,Drimmel2000,Siebert2011,Garcia2014,Gaia2018_KatzMW}, nor what is the exact size and orientation angle of the central bar  \citep[e.g.][]{Bissantz2003,Pettitt2014,Li2016} -- thus making it difficult to pin-point and study the large-scale distribution of molecular gas in the Galaxy. 

Progress is also being made in the characterisation of the earliest phases of (high-mass) star formation \citep[e.g.][]{urquhart2013_methanol,urquhart2013_cornish,urquhart2014_atlas,csengeri2014,csengeri2017,Trafi2017,Elia2017,urquhart2018,Pitts2019}, but the role of large scale structures and the interplay between the various phases of the interstellar medium (ISM) are still not well constrained.
Key questions remain, that are also relevant to the study of star formation in external galaxies, such as: what role do the spiral arms play in the formation of molecular clouds and star formation, and what controls the star formation efficiency (SFE).
Observations of nearby spiral galaxies have revealed a tight correlation between dense molecular gas and enhancements of star formation activity within spiral arms \citep[e.g.][]{Leroy2017}.
Also in the Milky Way, it is clear that spiral arms are rich in molecular gas. For example, recent results from the THOR survey reveal an increase by a factor 6 of atomic to molecular gas ratio from the arms to inter-arm regions.
However, it is not clear whether this is due to the collection of molecular clouds that fall into their gravitational potential \citep[e.g.][]{Foyle2010}, or if the molecular gas forms within the spiral arms themselves. 
Furthermore, it is unclear if the enhanced star forming activity observed (\citealt{urquhart2014_rms}) is directly attributable to the presence of spiral arms or is simply the result of source crowding within the arms (\citealt{moore2012}).

The SFE is the result of a number of stages: the conversion of neutral gas to molecular clouds, then to dense, potentially star-forming clumps, and finally to proto-stars and young stellar objects. Each of these stages has its own conversion efficiency. Identifying the stage that is primarily affected by the environment could bring constraints on the dominant SF-regulating mechanism.
Some progress has been made recently to investigate the effects of environment on the SFE in our Galaxy, based on limited samples of objects \citep{eden2012, moore2012, longmore2013, ragan2018}.
In order to extend these studies to much larger samples, we have performed a large scale ($\sim$84~deg$^2$) spectroscopic survey of the inner Galactic disc: the SEDIGISM (Structure, Excitation and Dynamics of the Inner Galactic Interstellar Medium) survey. The spectroscopic data provide essential information on the distribution of interstellar matter along the line of sight, thus complementing the existing continuum surveys. These data allow us to achieve an unbiased view of the moderately dense ISM over a large fraction of the Galactic disc. 
The SEDIGISM survey covers a large portion of the fourth quadrant at high velocity and angular resolution and will make a significant contribution to our understanding of Galactic structure.

The survey has been described in \citet[][hereafter Paper\,I]{Schuller2017}. It consists of spectroscopic data covering the inner Galactic plane in the frequency range 217$-$221~GHz, which includes the \cco\ and \coo\ molecular lines, at 30 arcsec angular resolution. Thus, this survey complements the other spectroscopic surveys that have been previously mentioned. 
This is the first of three papers that describe the survey data and present the initial results.
In the present paper, we provide an overview of the survey data and a first look at the connection between the molecular gas and large scale structural features of the Galaxy (we will refer to this as Paper\,II).
In the accompanying papers, we present a catalogue of giant molecular clouds (GMCs) and investigate their properties with respect to their star formation activity and their Galactic distribution (\adcprep; hereafter Paper\,III);
and we investigate the dense gas fraction and star formation efficiency as a function of Galactic position (Urquhart et al.\ submitted; hereafter Paper\,IV). 

The structure of this paper is as follows: we describe the SEDIGISM observations and data quality in Sect.\,\ref{sec:data}. We present the large scale distribution of $^{13}$CO and C$^{18}$O in Sect.\,\ref{sec:global_distribution} and investigate the association between molecular gas and spiral arms. We discuss some interesting regions in Sect.\,\ref{sect:interesting_regions}, and demonstrate the usability of other molecular transitions within the spectral range covered by the data for scientific exploitation in Sect.\,\ref{sect:exotic_lines}. Finally, we summarise our conclusions in Sect.\,\ref{sec-perspective}.

\section{The survey data}
\label{sec:data}

\subsection{Observations}
\label{sec-obs}

\begin{table}
\begin{minipage}{\linewidth}
\begin{center}
\caption{Summary of the APEX observational parameters.}\label{tab:obs_parameters}
\begin{tabular}{lc}
\hline
Parameter & Value \\
\hline
Galactic longitude range      &      -60\degr\ $< \ell <$  18\degr\ \\
 & and  29\degr\ $< \ell <$  31\degr\\
Galactic latitude range$^a$      &      $-$0.5\degr\  $< b <$ 0.5\degr\ \\
Instrument & SHeFI (APEX-1) \\
Frequency & 217--221\,GHz \\ % previously 230.538????
Bandwidth & 4\,GHz \\
Angular resolution & 30\arcsec \\
Velocity resolution & 0.1\,\kms\ \\
Smoothed velocity resolution & 0.25\,\kms\ \\
Mean noise (\tmb)$^b$ & $\sim$0.8$-$1.0\,K \\
Main beam efficiency ($\eta_{\rm mb}$)$^c$  &  0.75\\
\hline
\end{tabular}
\end{center}
\end{minipage}
\footnotesize{$^a$ The latitude range was extended to $-$1\degr\  $< b <$ 1\degr\ towards the Central Molecular Zone, to $b< -$0.75\degr\ towards the Nessie filament at $\ell\sim338\degr$, and to $b< +0.75$\degr towards the RCW~120 region at $\ell\sim348$\degr.}\\
\footnotesize{$^b$ Per 0.25\,\kms\ channel.} \\
\footnotesize{$^c$ http://www.apex-telescope.org/telescope/efficiency/index.php.}

\end{table}

Observations were done with the 12~m diameter Atacama Pathfinder Experiment telescope (APEX, \citealp{Guesten2006}), located at 5100~m altitude on Llano de Chajnantor, in Chile.
The observational setup and observing strategy have been described in Paper\,I and the key observational parameters are summarised in Table\,\ref{tab:obs_parameters}; here we provide a brief overview of the most important features of this survey.

\begin{figure*}
	\includegraphics[width=\textwidth]{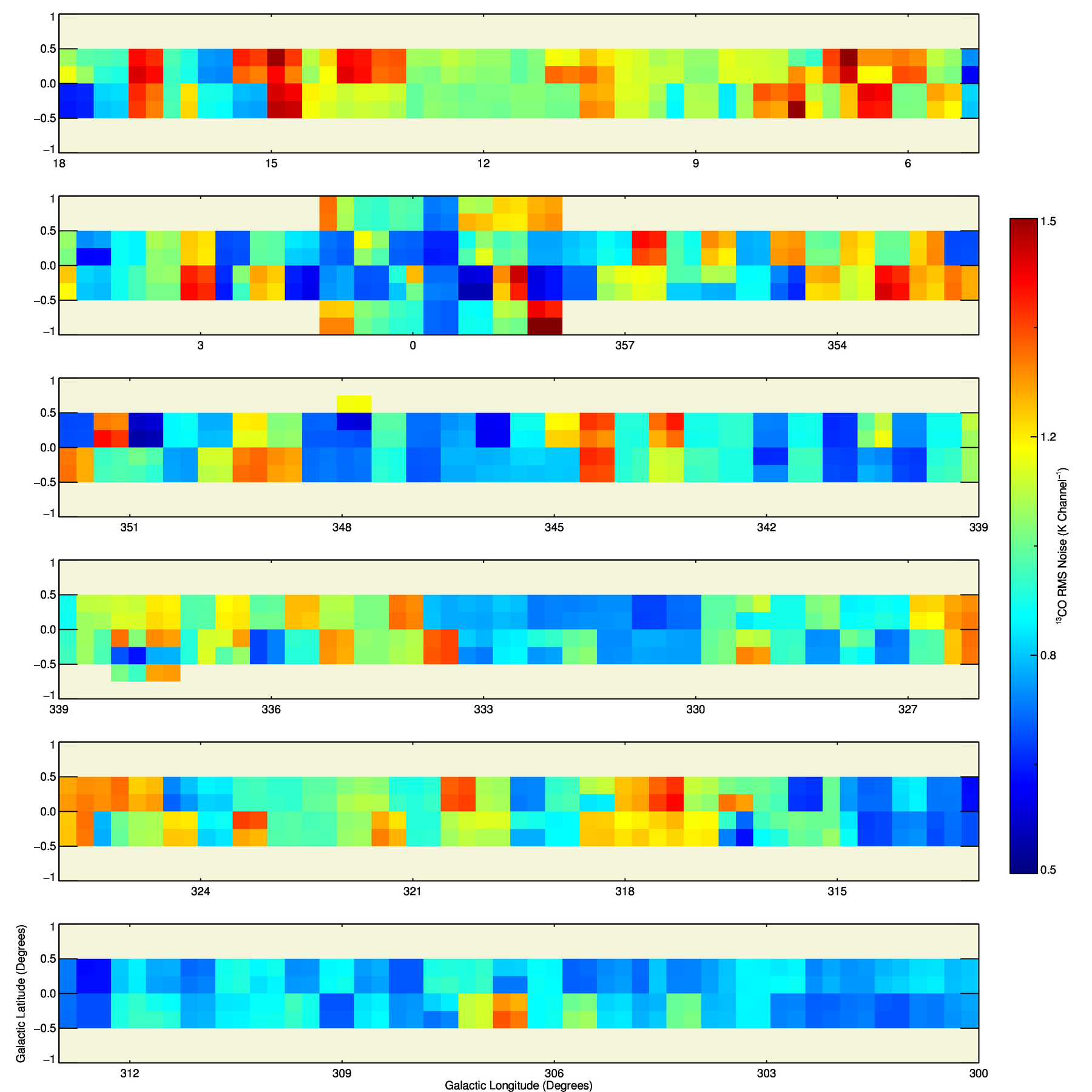}
   \caption{Spatial distribution of the local noise $\sigma_{\rm rms}$ over the survey area (see text for details).
   The median values computed in 0.25$\times$0.25~deg$^2$ sub-cubes are shown here in colours (see colour scale on the right)}
    \label{fig:allrms}
\end{figure*}

The prime target lines are \cco\ and \coo\, but the 4\,GHz instantaneous bandwidth also includes a number of transitions from other species (H$_2$CO, CH$_3$OH, SO, SO$_2$, HNCO, HC$_3$N, SiO).
The observations have been carried out in tiles of $0.25\degr \times 0.50\degr$ using position-switching in the on-the-fly mapping mode. Each position in the survey is covered by at least two
maps observed in orthogonal scanning directions, along galactic longitude and latitude. The $0.25\degr \times 0.50\degr$ tiles oriented along $\ell$ or $b$ were sometimes observed under different conditions. As a result of this plaiting, only $0.25\degr \times 0.25\degr$ sub-cubes were observed with roughly constant conditions and show a uniform noise level, as visible in Fig.~\ref{fig:allrms}.
Some fields were observed in $0.5\degr \times 0.5\degr$ maps at the beginning of the survey, also with two orthogonal scanning directions.

The reference positions for each field were selected to be $\pm1.5\degr$ off the Galactic mid-plane to avoid contamination, and while this was sufficient to ensure the \coo\ data were clean, this was not always the case for the brighter \cco\ transition.
Therefore, we have systematically performed pointed observations towards the references points, using an off position further from the Galactic plane.
More details can be found in Appendix~\ref{app:refspectra} and in Table~\ref{tab:refspectra}\footnote{Only a small portion of the data is provided here.
The full table is only available in the online version of this article https://doi.org/10.1093/mnras/staa2369}.

The full survey coverage is $\sim$84~deg$^2$ (cf.\ Fig.\,\ref{fig:allrms}): the main part of the survey, as described in Paper\,I, covers 300\degr $\leq$ $\ell$ $\leq$ +18\degr, with $\vert b \vert$ $\leq$ 0.5\degr\ (78~deg$^2$). Because additional observing time was available for this project in 2016 and 2017, we were able to slightly increase the survey coverage.
We have extended the coverage in latitude to $\vert b \vert$ $\leq$ 1\degr\ around the Galactic Centre (358\degr $\leq$ $\ell$ $\leq$ +1.5\degr) and mapped a 2~deg$^2$ region covering the extreme star forming region W43 (+29\degr $\leq$ $\ell$ $\leq$ +31\degr, with $\vert b \vert$ $\leq$ 0.5\degr).
We have also increased the latitude coverage up to +0.75\degr\ at $\ell = $ 348\degr\ to cover RCW~120,  and down to $-$0.75\degr\ for 338\degr\ $\leq \ell \leq$ 339\degr\ to improve the coverage of the Nessie giant filament \citep{jackson2010}.
The data taken towards W43 allow for a comparison with existing surveys in the northern hemisphere, such as the HERO survey performed with the IRAM 30-m telescope in the same spectral lines (\citealt{Carlhoff2013}). Here we present all the data that have been taken with APEX for this survey between 2013 and 2017.

%%%%%%%% Figure: map of median(rms)

\subsection{Data reduction}

The data provided by the APEX telescope consist of spectra calibrated in antenna temperature scale ($T^{\star}_{\rm A}$), written in files readable by the CLASS software from the GILDAS package\footnote{\url{http:/www.iram.fr/IRAMFR/GILDAS/}}. 
We have developed a dedicated pipeline in GILDAS/CLASS, which consists of standard data reduction steps, such as conversion to \tmb\ scale (i.e. \tmb\ $= T^*_{\rm A}/\eta_{\rm mb}$), spectral resampling, removing a spectral baseline, and gridding the spectra onto a data cube. In particular, a critical step consists in an automatic detection of emission features in order to define the windows to be masked when subtracting baselines; this is described in detail in Paper\,I.

%%%%%%%% Figure: rms histogram
\begin{figure}
	\includegraphics[width=0.49\textwidth]{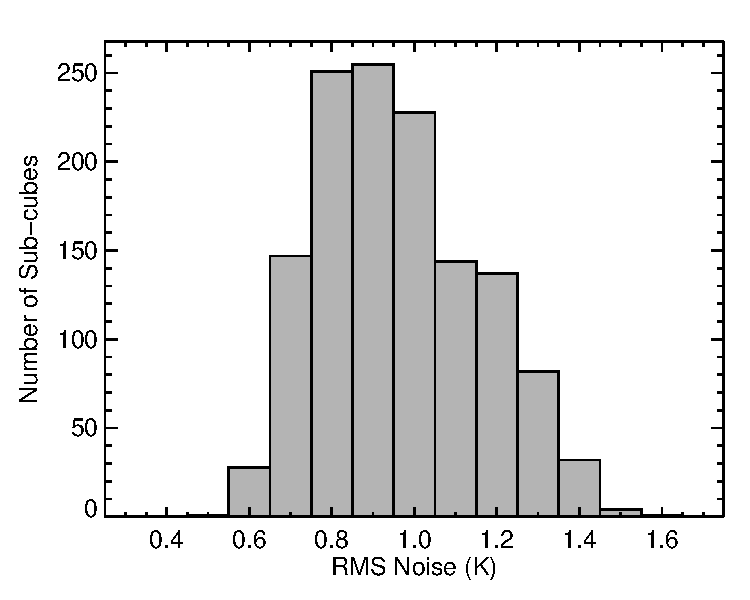}
   \caption{Distribution of the median values of the r.m.s.\ noise in 0.25$\times$0.25~deg$^2$ sub-cubes, shown in 0.1\,K (\tmb) bins.
   The mean and median values are both approximately 0.95\,K.}
  \label{fig:histo_rms}
\end{figure}
%%%%%%%% %%%%%%%% %%%%%%%% %%%%%%%% 

As previously mentioned, in some cases the initial off-position was found to contain emission that would contaminate our maps and so these were checked against a more distant position, one degree further away perpendicular to the Galactic plane, and reduced independently. Emission in the reference position appears as an absorption feature that is constant over the extent of a given map. When coinciding with the velocity range of real emission in the map, it leads to underestimating the gas column density, and could also impact the observed velocity pattern of the emission. Where the reference position has been found to show emission, we corrected for this by adding the spectrum measured for the reference position to each spectrum in the map.
This operation is illustrated in Fig.~\ref{fig:ex_correction} in Appendix~\ref{app:refspectra}.
Since the rms noise measured on these spectra was typically around 0.07~K (see Table~\ref{tab:refspectra}), this operation does not significantly increase the noise in the map data.

The pipeline used for the current first data release (DR1) is, therefore, almost identical to the procedure described in Paper\,I. However, an additional, non-standard data processing step was necessary; we noticed that when an on-the-fly map and the corresponding reference position were observed on different days, the Doppler corrections applied to the reference position and to the map data could differ by up to three velocity channels (0.75\,\kms). Therefore, to properly account for the absorption feature in the spectra where necessary, we computed this difference between the Doppler corrections computed for the map centre and for the reference position observed at a different time. After shifting the observation of the reference spectrum accordingly, we then added the modified reference spectrum to each position of the map. This step was only necessary where the reference position has been found to show significant emission (see Table\,3 for details).

\begin{table}
\centering
\caption{Fields with issues.}
\label{tab:issues}
\begin{tabular}{ll}
\hline
Field & Comments \\
\hline
304.25$-$0.25 & Baseline ripples \\
332.25$+$0.25 & Absorption feature at $-23$\,\kms\ \\
            & from the reference position \\
332.75$+$0.25 & Absorption features at $-23$\,\kms\ \\
            & and $-31$\,\kms\ from the reference position \\
336.25$-$0.25 & Absorption feature at $-20$\,\kms\ \\
            & from the reference position \\
352.25$+$0.25 & Absorption feature at $-6$\,\kms\ \\
            & from the reference position \\
\hline
\end{tabular}
\end{table}

%%%%%%%%%%%
\begin{figure*}
  \includegraphics[height=\textwidth, angle=0]{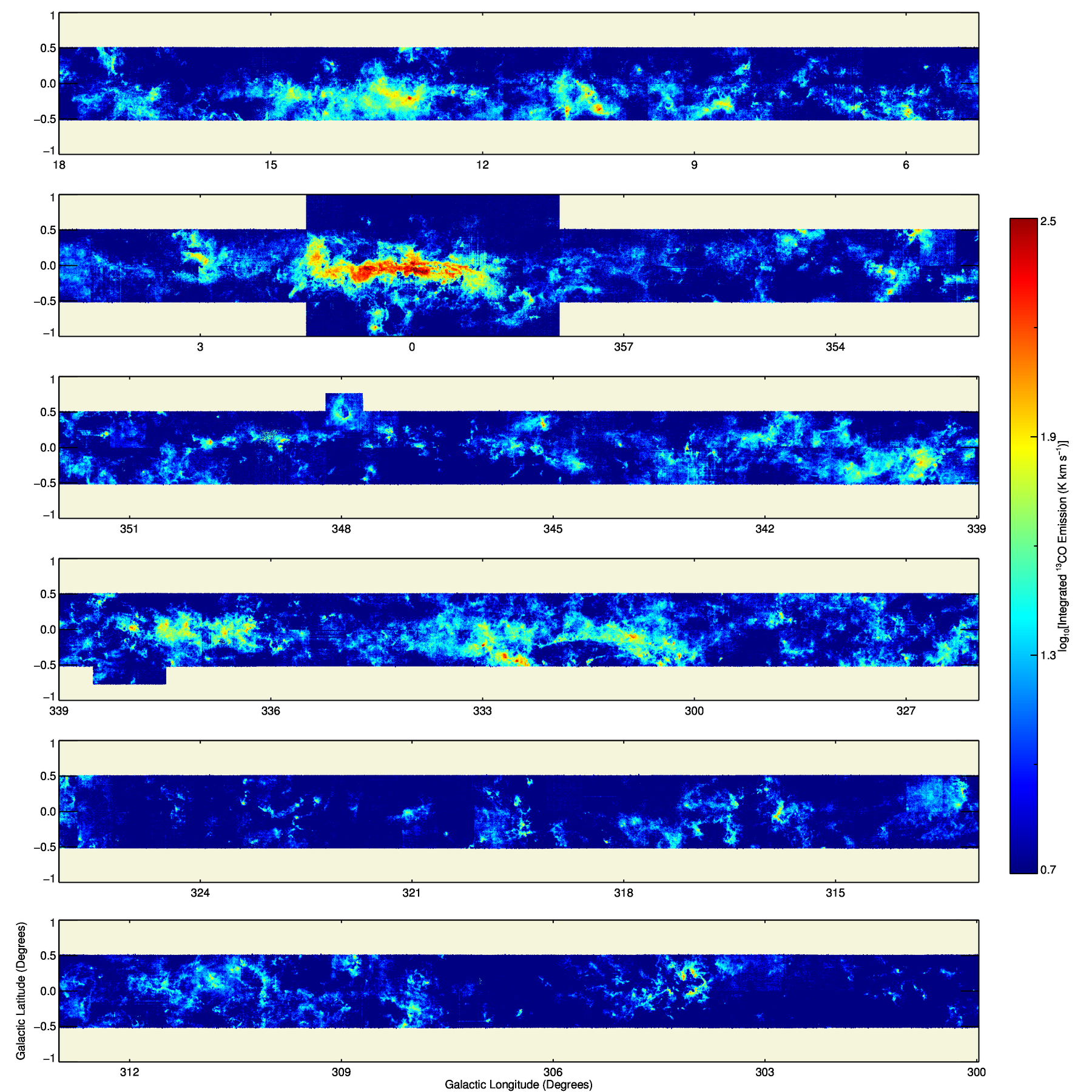}
  \caption{Integrated emission maps for the full survey (except for W43), showing the \cco\ spectral cubes integrated over the full $\pm$200\,\kms\ range of the data.
  }
 \label{fig:integ_map}
\end{figure*}

\subsection{Data quality}
\label{sec:quality}

The output of the pipeline is a set of data-cubes calibrated to the \tmb\ scale, centred on the most relevant spectral lines possibly detected in the band-pass, and projected on 9.5$''$ pixels. The velocity resolution in the final cubes is 0.25\,\kms\ for the \cco\ and \coo\ lines, and 0.5\,\kms\ for all the other transitions: H$_2$CO(3$_{0,3}-2_{0,2}$) and ($3_{2,1}-2_{2,0}$), HC$_3$N($24-23$), SO($6_5-5_4$), SiO($5-4$), HNCO(10$_{0,10}-9_{0,9}$), and CH$_3$OH (4$_{2,0}-3_{1,0}$).

Due to variations in weather conditions, the elevation of the telescope during observations, and the performance of the instrument, the local noise level $\sigma_{\rm rms}$ varies slightly over the survey area, as illustrated in Fig.\,\ref{fig:allrms}.
For each pixel in each \cco\ data cube, $\sigma_{\rm rms}$ is computed as the standard deviation of the signal in the first 50 spectral channels (i.e. $-200$\,\kms\ $ < v_{\rm lsr} < -187.5$\,\kms), as this part of the velocity range is almost always devoid of emission, except possibly for some small regions near the Galactic centre.
In Fig.~\ref{fig:histo_rms} we show the distribution of the median noise values for all of the individual $0.25\degr \times 0.25\degr$ sub-cubes, which shows that the range of noise values is between 0.5--1.6\,K (\tmb), with most fields (61 per cent) showing a noise below 1.0\,K; the median and mean values are 0.94 and 0.96\,K respectively.
While the sensitivity achieved is sufficient to map the distribution of the relatively bright \cco\ line emission, we are only able to detect \coo\ towards the densest regions. Transitions from other species are likely to be detected only towards the brightest and most compact dense cores.

Another issue worth highlighting is that subtracting baselines in regions with bright, broad emission lines is known to be very arduous and error-prone, particularly when the baseline exhibits variations over a velocity range comparable to the line-width. The central region of the Galaxy is clearly the most extreme case in that respect, and the data for this region should be used with particular caution.
In addition, there are a few other regions that are affected by baseline ripples and absorption features that we have not been able to fully remove by adding the spectrum of the corresponding reference position. We identify these regions in Table\,\ref{tab:issues} and recommend extreme caution when using data for these fields.
However, this issue only emerges when averaging spectra over large regions; correction from the reference position to the individual spectra does not show such artefacts.
Finally, the \coo\ data shows a spike at \vlsr\,$\sim-48$\,\kms\ that appears in a single channel with a varying intensity over all fields. We have removed this spike from the individual spectra using a sigma-clipping method, however, an artefact may still show up when averaging the spectra over large areas.

In order to check the consistency of our calibration and the data quality, we have compared the SEDIGISM data for the W43 complex with the W43-HERO survey \citep{Carlhoff2013}, which covered a good fraction of the 2 deg$^2$ field centred at $\ell =$ 30\degr\ in the same transitions as SEDIGISM, \cco\ and \coo, with the IRAM 30~m telescope.
The results from this comparison show no systematic differences between the two data sets for the \cco\ line and indicate that the calibration is consistent between the two surveys, although the distribution of the $^{13}$CO emission reveals some minor differences towards the brightest areas.
Since the C$^{18}$O SEDIGISM data is only tracing the most compact, dense clumps but is not sensitive to the more diffuse, extended material, no statistical comparison is possible (see Appendix\,\ref{A:sed_hero_comp} for more details).

\subsection{Public data release}

The reduced data, processed with the current version of the dedicated pipeline, is now available to the community. This first public data release (DR1) consists of 78 cubes in \cco\ and 78 cubes in \coo, where each cube covers 2\degr\ in longitude over $\pm$0.5\degr\ in latitude (or more in the few regions with extended $b$ coverage, see \S~\ref{sec-obs}). Cubes are separated by 1\degr\ in longitude, providing 1\degr\ overlap between adjacent cubes. The pixel size is 9.5$''$, the velocity resolution is 0.25\,\kms, and the velocity range covers $-$200 to +200\,\kms. The DR1 data can be downloaded from a server hosted at the MPIfR in Bonn (Germany).\footnote{\url{https://sedigism.mpifr-bonn.mpg.de}}

The 4\,GHz of instantaneous bandwidth of the spectral tuning used to observe SEDIGISM includes several transitions from other molecules (see Table\,1 in Paper\,I). In particular, six lines are detected when spatially averaging the data corresponding to the brightest emission regions in \cco; this will be discussed in more detail in Sect.\,\ref{sect:exotic_lines}.
Therefore, we also provide data cubes covering these transitions as part of the DR1, where the spectra have been smoothed to 0.5\,\kms\ velocity resolution in order to increase the signal-to-noise ratio. The velocity range also covers from $-$200 to +200\,\kms\ for all lines, except for the SiO (5--4) transition, which is located near the edge of the spectral set-up so that only the $-$200 to +150\,\kms\ velocity range is available.

In Paper\,III, we have extracted $\sim$11,000 molecular clouds and complexes from the \cco\ data from this data release, using the Spectral Clustering for Interstellar Molecular Emission Segmentation \citep[SCIMES,][]{ref-scimes} algorithm. The catalogue with the distance estimates and the derived physical properties for all clouds, as derived in Paper\,III, as well as the masks representing these clouds in the \cco\ cubes are also made publicly available alongside the main data release we present here.

Some work is still ongoing aimed at improving the data quality and at solving known issues, such as artefacts for some transitions or a proper estimate of the baselines in regions with complex, extended emission. We plan to provide data cubes with improved quality as part of future data releases.

%%%%%%%%%%%%%%%%%%%%%%%%%%%%%%%%%%%%%%%%%%%%%%%%%%%%%%%%%%%%%%%%%%
%%%
\section{Global distribution of molecular gas}
\label{sec:global_distribution}

In this section, we use the \cco\ and \coo\ data to constrain the Galactic structure on the largest scale. Some peculiar regions will be discussed in more detail in Sect.~\ref{sect:interesting_regions}, while compact objects and individual molecular clouds are the topic of subsequent papers.

\subsection{Integrated emission}

We show maps of the $^{13}$CO emission integrated over the $\pm$200~\kms\ velocity range for the full survey in  Fig.\,\ref{fig:integ_map}. These maps reveal that the emission from molecular gas is extended over much of the inner part of the survey region ($330\degr < \ell <  15\degr$) but becomes much more patchy at larger angular distances from the Galactic centre. The brightest emission is associated with the Central Molecular Zone (CMZ; \citealt{morris1996}), that extends over $\vert \ell \vert \leq 1.5\degr$ or a radius of $\sim$200~pc.
Outside of the Galactic centre the brightest emission is concentrated in distinct regions, all associated with prominent star forming regions.

Our latitude coverage is rather narrow: only $\pm$0.5\degr\ in most directions.
By comparing with other surveys with larger latitude coverage like  ThrUMMS or ATLASGAL, it is clear that we miss a number of molecular clouds and complexes, especially the nearby ones \citep[see also][regarding nearby clouds in the outer Galaxy]{Alves2020}. Also a few known complexes up to $\sim$3~kpc are not included, e.g.\ NGC~6334 and NGC~6357.
Another exception is for gas lying towards the longitude range of $300\degr < \ell <318\degr$, and a Galactocentric distance of 8\,kpc and beyond, where the Galactic plane descends below a latitude of $b < -0.5\degr$ due to the Galactic warp \citep[e.g.][]{Chen2019,Romero-Gomez2019}, making this area of the Galaxy not well covered by our survey.
However, the area covered by SEDIGISM still encompasses the vast majority of the molecular gas in the inner Galaxy; \citet{rigby2016} also concluded from a comparison of the CHIMPS data (with the same latitude coverage as SEDIGISM) with the GRS survey ($|b|< 1\degr$) that the molecular line emission drops off quickly with distance from the mid-plane.

\subsection{Longitude-velocity distribution}

\subsubsection{A four-spiral-arm model}

\begin{figure}
  \includegraphics[width=0.49\textwidth, trim= 0cm 0cm 0cm 0cm, clip]{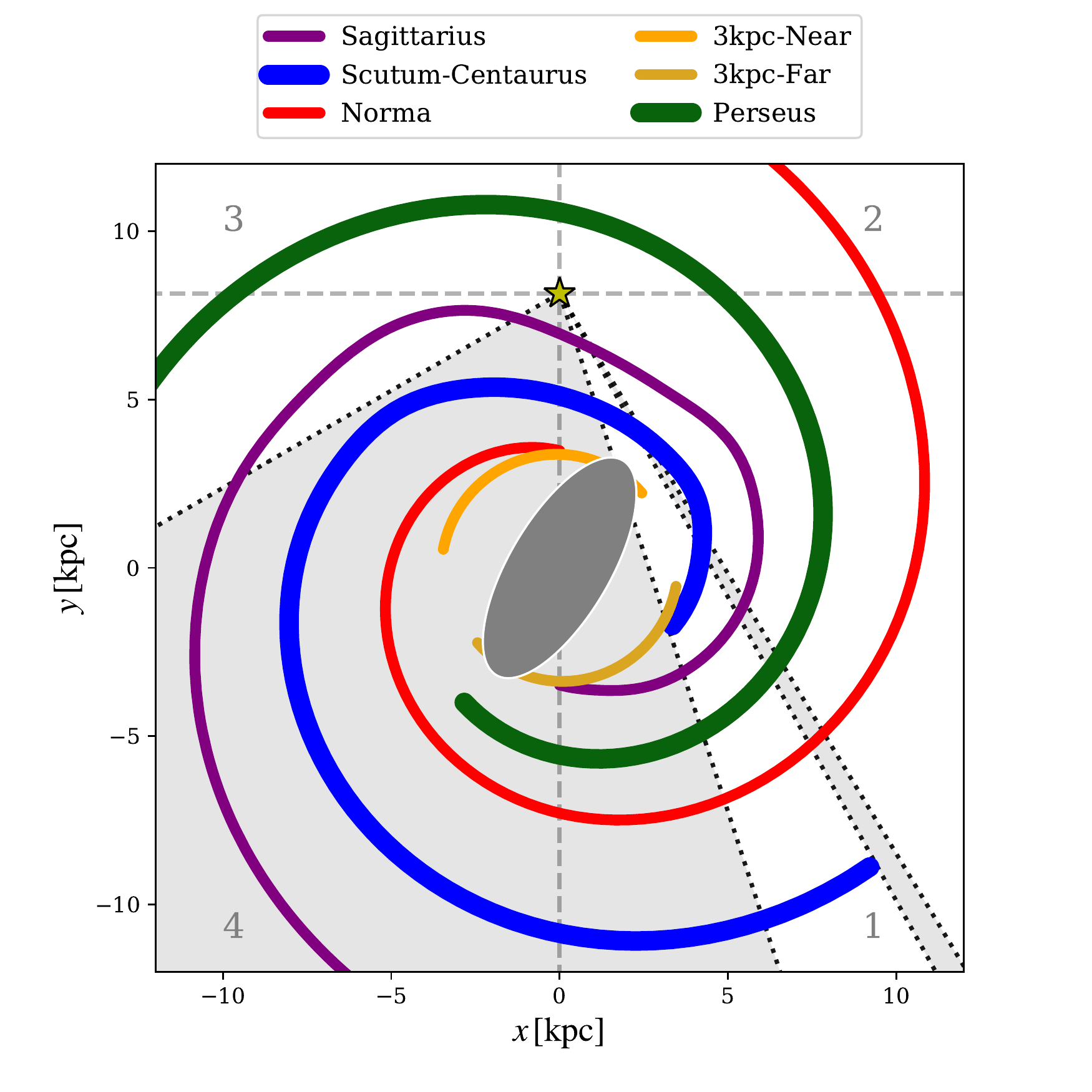}
  \vspace{-5mm}
  \caption{Schematic showing the loci of the spiral arms according to the model by \citet{taylor1993} and updated by \citet{cordes2004}, with an additional bisymmetric pair of arm segments added to represent the 3~kpc arms. The grey shaded areas indicate the regions covered by the SEDIGISM survey. The star shows the position of the Sun and the numbers identify the Galactic quadrants. The bar feature is merely illustrative and does not play a role in our analysis. The smaller slice in the first quadrant corresponds to the W43 region. }
    \label{fig:topdown_view}
\end{figure}

To facilitate the discussion that follows concerning the distribution of molecular material in the Milky Way, Fig.\,\ref{fig:topdown_view} presents a schematic plot of the Galaxy as seen from the northern Galactic pole that includes the spiral arms and the Galactic bar. We have chosen to use the spiral arm loci derived by \citet{taylor1993} and updated by \citet{cordes2004} as these have been determined independently of the distribution of molecular gas unlike the loci used in the recent work by \citet{reid2019}, which have been fitted by-hand to the $^{12}$CO\,(1$-$0) $\ell v$-map of \citet{dame2001}. Moreover, the model from \citet{reid2019} is poorly constrained in the fourth quadrant due to a lack of reliable maser parallax distances.
A single  galactic bar is shown for illustrative purposes, with orientation and length scales in line with contemporary measurements \citep{bland-hawthorn2016}.
The near and far 3~kpc arms are not included in the \citet{cordes2004} model and have been added in as small, 2-fold symmetrical arm segments \citep{Dame2008} with pitch angles of 1.5$^\circ$ expanding radially at a velocity of 55\,\kms, with the near 3~kpc arm aligning with the arm segment from \citet{bronfman2000} (the exact nature of these features is still somewhat unknown, see \citealt{green2011}). There is some evidence that the far-3~kpc arm is expanding slightly faster, but this is inconsequential for our analysis as there is effectively no emission seen in this region in the SEDIGISM data, due to the limited sensitivity.
Fig.\,\ref{fig:topdown_view} shows that the SEDIGISM survey (indicated by the grey shading) covers large parts of three of the main spiral arms (\nor, \sag, and \scu\ arms), and almost all of the 3~kpc arms, thus allowing us to refine our understanding of the structure of the Galaxy. 

\begin{figure*}
\includegraphics[width=0.9\textwidth]{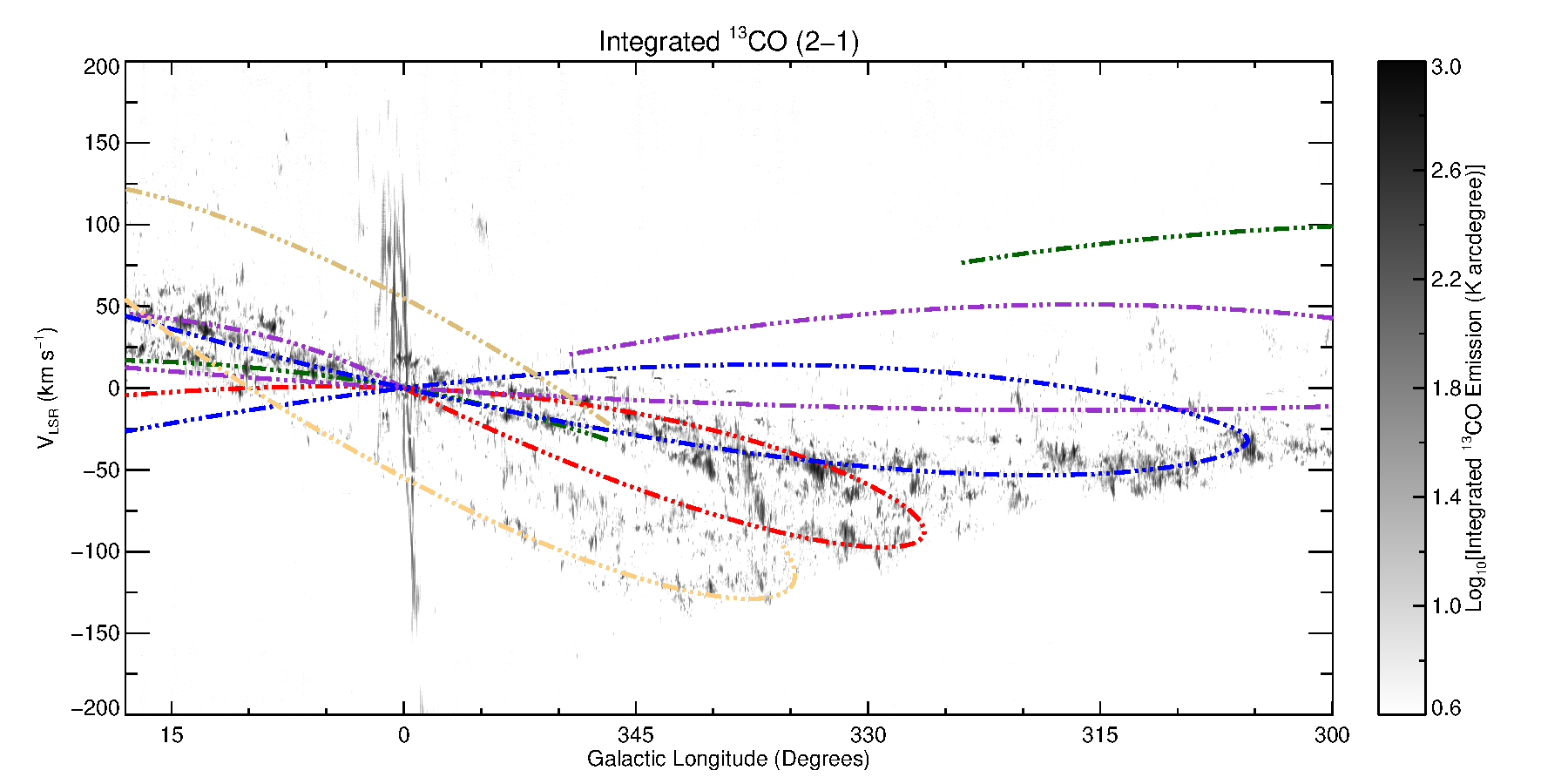}
\includegraphics[width=0.9\textwidth]{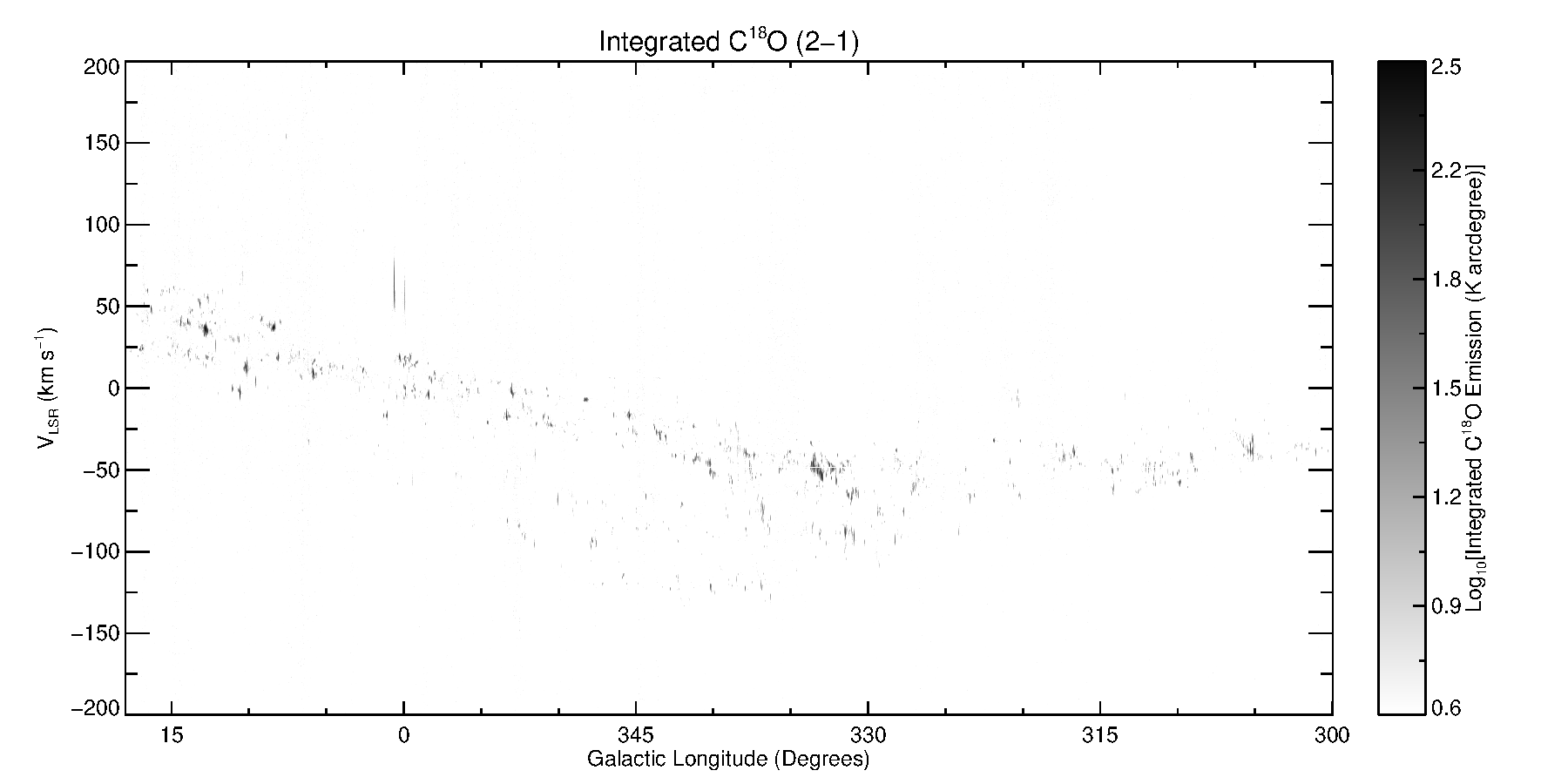}
\includegraphics[width=0.9\textwidth]{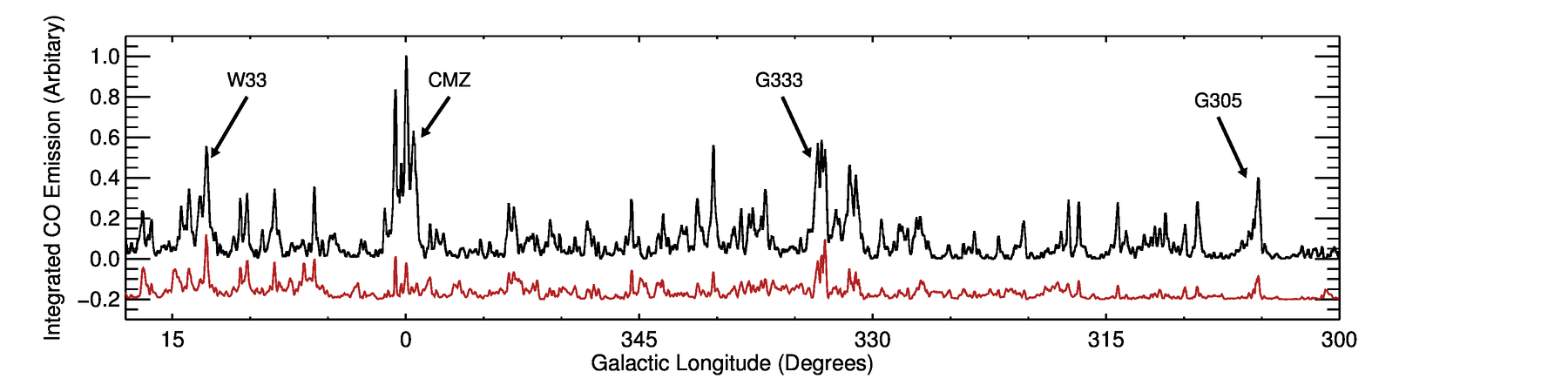}
  \caption{Galactic longitude-velocity distribution of the SEDIGISM survey between $300\degr < \ell < 18\degr$. The greyscale image shows the distribution of molecular gas as traced by the integrated \cco\ and \coo\ emission (upper and middle panels). To emphasis the weaker extended emission we have used a log scale and have masked the emission below 3$\sigma$. The intensity in \tmb\ scale has been integrated over the $\pm$0.5\degr\ range in Galactic latitude. The location of the spiral arms are shown as curved dotted-dashed lines, coloured to identify the individual arms; colours are as shown in Fig.\,\ref{fig:topdown_view}.
  For the \coo\ line, the values of a horizontal row of three pixels centred on $-48.5$\,\kms\ have been set to zero due to the presence of a spike that appears at this velocity when large areas are integrated together (for more details see Sect.\,\ref{sec:data}).   Lower panel: Integrated \cco\ and \coo\ intensity as a function of Galactic longitude (black and red respectively). The intensities have been integrated over the $\pm$200\,\kms\ in \vlsr\ and the $\pm$0.5\degr\ range in latitude for each longitude. The flux scale has been normalised to the peak intensity of the \cco\ emission. The \coo\ spectrum has been multiplied by 5 and an offset of $-0.2$ has been applied to make the profile clearer. }
    \label{fig:lv_plots}
\end{figure*}

\subsubsection{SEDIGISM $\ell v$ maps}

 In the upper and middle panels of Fig.\,\ref{fig:lv_plots} we present a longitude-velocity map of the \cco\ and \coo\ transitions produced by integrating the emission between $|b| < 0.5$\degr.
 Only voxels above a 3$\sigma_{\rm rms}$ threshold were considered to produce this $\ell v$ map, where the local noise $\sigma_{\rm rms}$ is estimated as discussed above (Sect.~\ref{sec:quality}).
While much of the complex emission seen in the $\ell b$ map (Fig.~\ref{fig:integ_map}) is the result of many giant molecular clouds being blended along our line of sight across the inner Galactic disc, we find that these clouds are well separated in velocity, making it easier to break down the emission into distinct molecular structures. It is clear from these maps that while the \cco\ is detected over a wide range of velocities, the \coo\ emission is much less extended and is likely to only be tracing dense clumps, which cannot be easily detected beyond a few kpc due to beam dilution. But even if the \coo\ is less useful for studies of large scale structures, it is indispensable for detailed studies of the physical properties of dense structures such as filaments (\citealt{mattern2018}) and clumps (Paper\,IV).
 
\begin{figure*}
  \includegraphics[width=0.995\textwidth,trim= 1cm 0 0 0, clip]{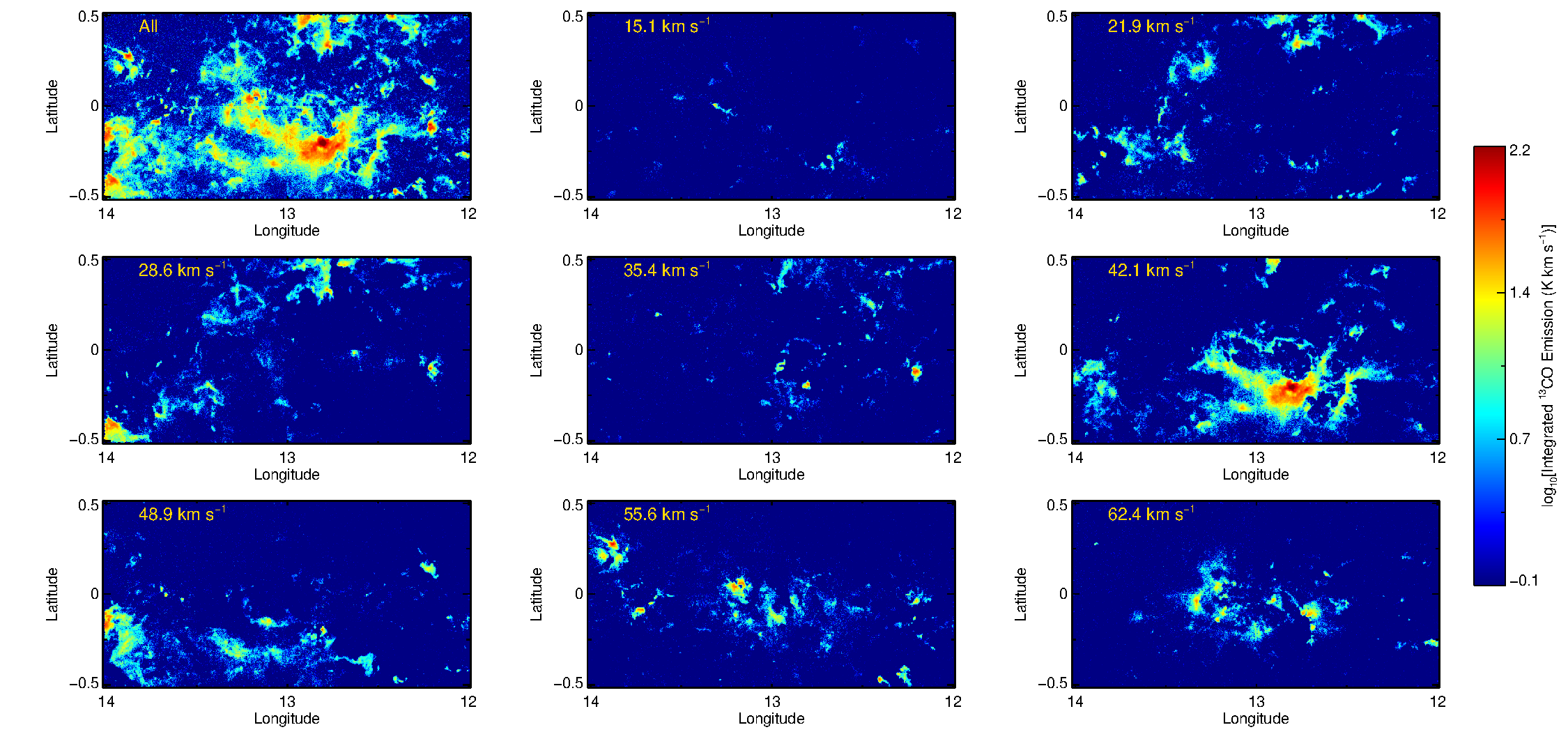}
  \caption{Channel maps of \cco\ of the $\ell =13\degr$ cube, which includes the high-mass star forming region W33. The upper-left panel shows the emission integrated over the full velocity range where emission is seen in this direction (0 to 60\,\kms), while the other panels show the emission integrated over velocity intervals of $\sim$6\,\kms. The central velocities over which the integration has been performed are given in the upper-left corners of each map. The emission in each map has been scaled to the brightest emission in each map. }
  \label{fig:channel_map}
\end{figure*}
 
On the \cco\ $\ell \varv$-map shown in Fig.\,\ref{fig:lv_plots}, we also overlay the spiral arm loci derived by \citet{taylor1993} and \citet{cordes2004}.
The $x$ and $y$ positions given by Taylor and Cordes have been converted to $\ell$ and $\varv$ using a three-component rotation curve (bulge + disc + dark halo)  tailored to the data of \citet{eilers2019}.
The shape of the rotation curve towards the Galactic centre is somewhat uncertain, and so we simply adopt a steeply rising bulge component matching that adopted in \citet[][Fig.~3]{eilers2019}. The \citet{reid2019} values for solar position and circular velocity at the orbit of the Sun, 8.15\,kpc and 236\,\kms, are used for projection into line-of-sight velocity, simply assuming pure circular rotation.
Comparing the CO emission to the loci of the spiral arms we find very good agreement outside the Galactic centre region ($\vert \ell \vert \leq 5\degr$).
The only significant region of emission that is not closely associated with a spiral arm is the CMZ, but this is known to have extreme non-circular velocities (this region is discussed in more detail in Sect.\,\ref{sect:interesting_regions}).
We also note that the majority of the CO emission is located within the solar circle (i.e. \vlsr\ $< 0$ for $\ell > 300$\degr\ and  \vlsr\ $> 0$ for $\ell > 0$\degr) and so there is very little emission seen towards the far parts of the \per, \sag, or \scu\ arm. This is likely the result of the sensitivity limit of the survey and beam dilution, and this means that probably we are only able to detect the most massive clouds outside the solar circle on the far-side of the Galaxy.

Our simple projection of arms into $\ell v$ space assumes purely circular motions for the primary arms, thus will not perfectly align with structures like the Norma Arm in the inner galaxy (appearing to move towards us with a \vlsr\ of roughly -30\,\kms, \citealt{Sanna2014}). The response of gas to spiral arms alone creates non-circular motions, forming some peculiar features towards the inner Galaxy seen in $\ell v$ space (e.g. \citealt{Gomez2004,Pettitt2015}).
Our modern, ${Gaia}$-era understanding of the Galactic bar also suggests slower pattern speeds than assumed in earlier works, which place corotation as far out as 6\,kpc \citep{Sanders2019,Bovy2019}. The ISM responds strongly to the motion of the bar out to corotation, and even as far as the more distant Outer Lindblad Resonance for certain models of bars \citep{Sormani2015,Pettitt2020}.
Any spiral arm-like features are thus inherently coupled to the bar within at least corotation and more sophisticated modelling is required to fully understand the kinematics of the gas.

In the lower panel of Fig.\,\ref{fig:lv_plots} we show the total \cco\ and \coo\ emission as a function of Galactic longitude; the \cco\ emission profile reveals a number of significant peaks, the most prominent of which is associated with the Galactic centre region. Many of the others are associated with well known star-forming complexes such as W33, G333 and G305; these are located at $\ell = 13\degr, 333\degr$ and 305\degr, respectively. The integrated \coo\ emission also reveals peaks that are correlated with the same star-forming regions indicating these regions have either higher optical depth and column densities than elsewhere in the Galactic plane, or they contain enough gas at high temperature to produce strong emission in the J=2--1 lines. 

Analysis of the dense gas traced by the ATLASGAL survey by \citet{urquhart2018} has shown that approximately 50 per cent of the current star formation in the disc of the inner Galaxy is taking place in a relatively small number of very active regions ($\sim$30). Indeed nearly all of the peaks seen in the lower panel of Fig.\,\ref{fig:lv_plots} are associated with one of these regions (see also \citealt{urquhart2014_rms}).
If we define all emission above a value of 10 per cent of the highest peak value (at $\ell \sim 0$) as being associated with a complex, we find that they are responsible for $\sim$70\,per\,cent of all the emission.
The $^{13}$CO emission traces the lower density diffuse gas in which the dense clumps are embedded and thus allow us to probe the structure, kinematics, and physical properties of these regions. This highlights the survey's ability to conduct detailed studies of the molecular gas associated with some of the most intense star formation regions in the Galaxy, and put them in a global setting with respect to the large scale structural features of the Galaxy.

In Fig.\,\ref{fig:channel_map} we show channel maps towards W33, a large complex representative of high-mass star forming regions in the Galactic disc.
According to maser parallax measurements, this complex is located in the \scu\ arm at a distance of 2.4\,kpc (\citealt{immer2013}). In the upper-left panel of this figure we show the integrated emission over the entire velocity range where emission is detected (0--60~\kms). Each of the subsequent panels shows a channel map where the emission has been integrated over 6\,\kms\ in velocity.
These maps reveal that the \cco\ emission seen towards this region consists of dense clumps, diffuse larger clouds and numerous filamentary structures spread out over 60\,\kms.
It is  worth noting that there is a wealth of intricate features that emerge in individual channel maps, but that do not appear or are washed out in the integrated intensity map.
This also implies that, even if most of the molecular gas is associated with a few major complexes as discussed above, there are plenty of other smaller features detected in the SEDIGISM data that are not associated with known complexes.

\subsection{Correlation between molecular gas and spiral arms}

The $\ell v$-map presented in Fig.\,\ref{fig:lv_plots} clearly shows that the molecular gas is broadly correlated with the spiral arms. To properly map the distribution of the molecular gas across the disc requires determining distances, which is beyond the scope of the current paper but is discussed in the accompanying paper by Duarte-Cabral et al. (Paper\,III).
However, it is possible to examine the intensity distribution as a function of the Galactocentric distance.  This is accomplished by calculating the kinematic distance for each pixel in the $\ell v$-map above a 3$\sigma_{\rm rms}$ threshold using the three-component rotation curve of \citet{eilers2019} and the \citet{reid2019} values for solar position and velocity, 8.15\,kpc and 236\,\kms\ (as described in Sect.\,\ref{sec:global_distribution}). Although this produces two distances for sources located within the solar circle (i.e. with a Galactocentric radius $R_{\rm gc}$ $< 8.15$\,kpc) equally spaced on either side of the tangent distance (referred to as the near and far distances) it provides a unique distance from the Galactic Centre which, therefore, allows us to investigate the distribution of the integrated intensity as a function of Galactocentric distance.
Given that the spiral structure is different in the 1$^{\rm{st}}$ and 4$^{\rm{th}}$ quadrants, and that the SEDIGISM survey has only covered a small portion of the 1$^{\rm{st}}$ quadrant, we have restricted this analysis to the 4$^{\rm{th}}$ quadrant. We have also excluded the Galactic Centre region ($|\ell| < 10\degr$) as kinematic distances are unreliable in this part of the Galaxy. 
Even outside this region, this approach cannot provide very accurate distances because of non-circular motions that deviate from the rotation curve, but it allows us to roughly estimate the fraction of molecular gas that is associated with the spiral arms.

\begin{figure}
  \includegraphics[width=0.49\textwidth]{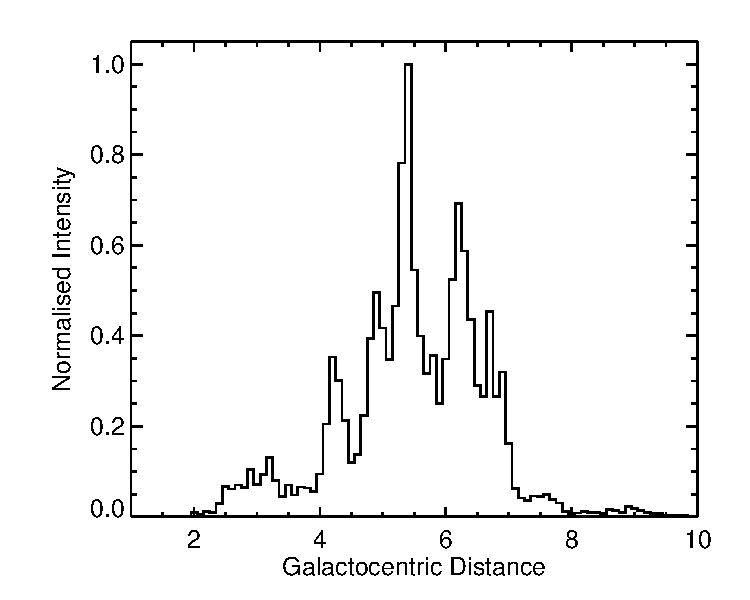}
  \caption{Integrated \cco\ intensity as a function of Galactocentric distance. The emission has been integrated over intervals of 0.1\,kpc and only includes longitudes between $300\degr < \ell < 350\degr$. }
  \label{fig:intensity_fn_RGC}
\end{figure}

In Fig.\,\ref{fig:intensity_fn_RGC}, we show the integrated \cco\ intensity as a function of Galactocentric distance, normalised to the peak of the distribution. This plot shows that the emission from the molecular gas is highly structured with strong peaks seen at approximately 4.25, 5, 5.5 and 6.5\,kpc; the first of these roughly corresponds to the tangent with the Norma arm, the second and third correspond to the far side of the long-bar where it intersects with the \per\ arm (\citealt{bland-hawthorn2016}), and the fourth with the tangent of the \scu\ arm. The vast majority of the emission is contained between 4 and 7.5\,kpc.
The lack of emission below 2\,kpc is due to the restricted longitude range selected for this analysis, while the lack of emission at distances greater than 8\,kpc likely reflects the poor sensitivity to molecular material on the far-side of the solar circle and beyond, where beam dilution certainly plays an important role.
This thick emission zone is analogous to the thick ring of material seen in the 1$^{\rm{st}}$ quadrant (often referred to as the 5\,kpc molecular ring), but as pointed out by \citet{jackson2006}, is likely to arise from a complicated combination of column density and velocity fields and may not actually represent a real ringlike structure \citep[see also][]{DB2012}. The highly structured nature of our \cco\ emission further lends support for a 4-arm model of the Galaxy (\citealt{urquhart2014_rms}), which can nevertheless co-exist with a ringlike structure.

\begin{figure}
  \includegraphics[width=0.49\textwidth]{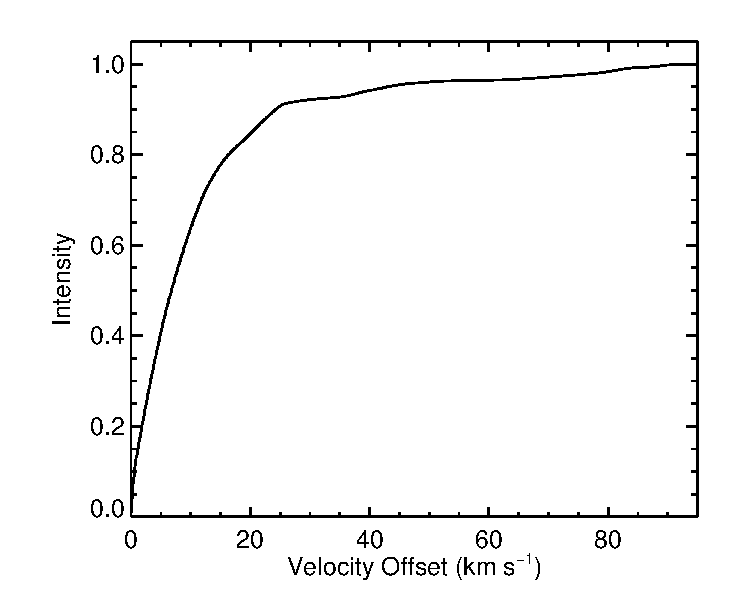}
  \caption{Cumulative integrated \cco\ intensity as a function of velocity offset from the nearest spiral arm for longitudes between $300\degr < \ell < 350\degr$. The velocity offsets have been calculated by finding the minimum velocity difference to a spiral arm for each pixel in the $\ell v$ map with a flux above 3$\sigma$.}
  \label{fig:vlsr_offset_from_arms}
\end{figure}

In order to quantify what fraction of the total emission is associated with the spiral arms we have calculated the minimum offset from the arms for each pixel on the $\ell v$-map above 3$\sigma_{\rm rms}$.
In Fig.\,\ref{fig:vlsr_offset_from_arms} we show the cumulative distribution of the integrated \cco\ emission as a function of velocity offset from the spiral arm loci shown in the upper panel of Fig.\,\ref{fig:lv_plots}. When performing the matching of the pixels with the spiral arms we allowed for a variation of $\pm0.5$\degr\ in Galactic longitude as the spiral arm tangents are not well constrained. We consider pixels within $\Delta v < 10$\,\kms, which is of the order of the amplitude of streaming motions around the spiral arms ($\sim$7-10\,\kms; \citealt{burton1971,stark1989,reid2009}), to be associated with a spiral arm. This plot reveals that approximately 60 per cent of the molecular emission is closely associated with a spiral arm. This proportion is a little lower than the value of 80 per cent derived by \citet{urquhart2018} from a similar analysis of GRS clouds identified by \citet{rathborne2009}. However, as pointed out by \citet{roman2009} only approximately two-thirds of the emission in the GRS was accounted for in the source extraction with diffuse emission below the detection threshold accounting for the rest.
The strong correlation we have found between the $^{13}$CO emission and the spiral arms is consistent with the findings of \citet{roman2009} and \citet{rigby2016}. Nevertheless, this analysis also indicates that a significant amount of molecular gas (up to 40\,per\,cent) is located in the inter-arm regions.

\section{Noteworthy regions}
\label{sect:interesting_regions}

Although it is clear that the majority of the $^{13}$CO emission outside the CMZ is closely associated with the spiral arms, there are a number of interesting features seen in the $\ell v$ map that are worth discussing in some detail.

\subsection{Galactic Centre}

\begin{figure*}
\includegraphics[width=0.99\textwidth]{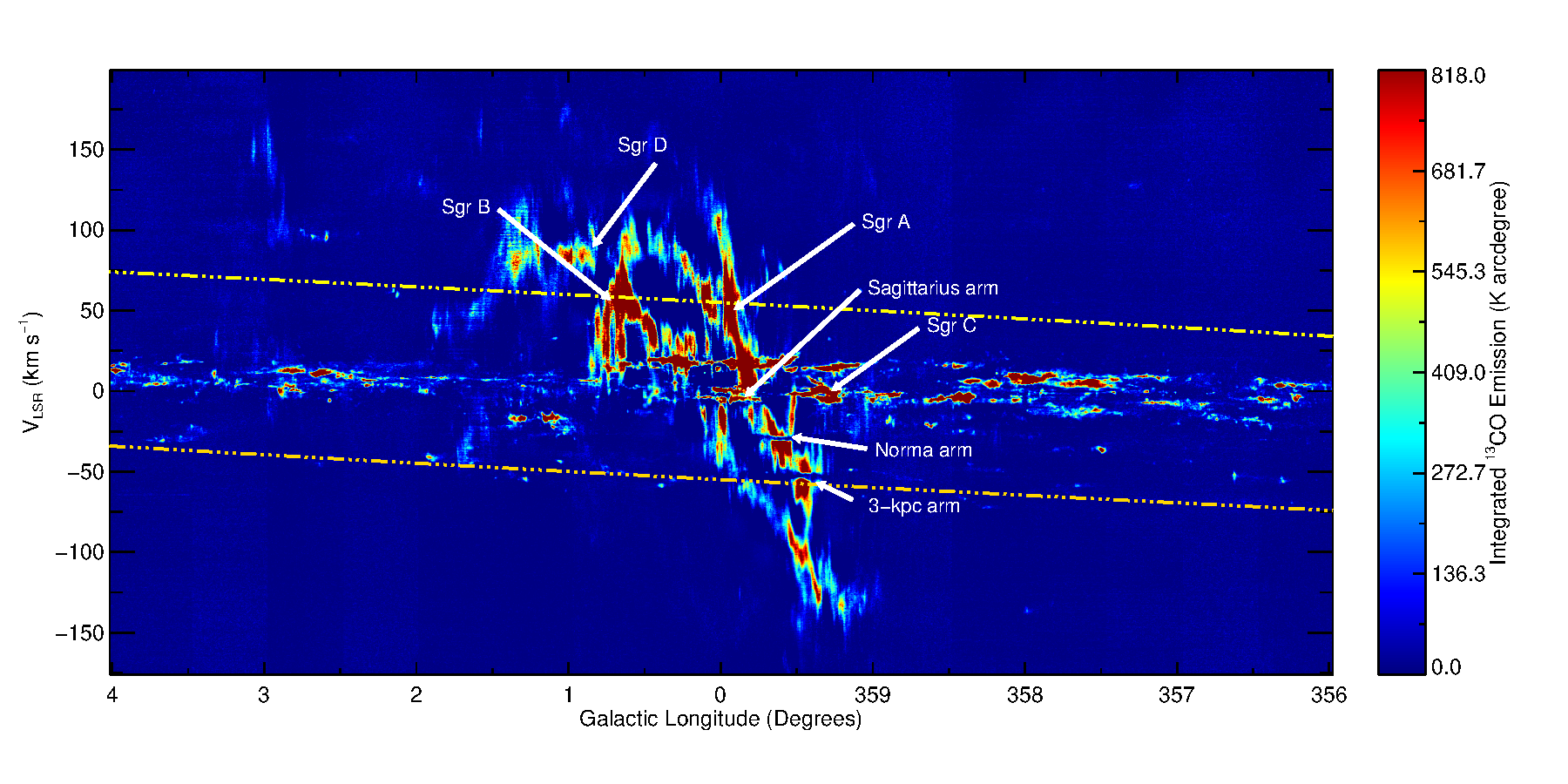}
  \caption{Longitude-velocity map of the Galactic centre region. The lower and upper yellow dashed-dotted lines show the loci of the near and far 3~kpc arms respectively (see text for details). On this map we label some of the more significant molecular clouds and absorption features that have been attributed to foreground spiral arms; some of these are discussed in the text.}
    \label{fig:gc_region}
\end{figure*}

In Fig.\,\ref{fig:gc_region} we show the $\ell v$-map of the Galactic Centre region. The emission in this region can be separated into two distinct types: a narrow horizontal strip of emission centred around \vlsr = 0 that stretches across the whole map, and a region of more complex emission between $359\degr < \ell < 2\degr$ and velocities $-$150\,\kms\ $<$ \vlsr $<$  150\,\kms.
The horizontal strip around \vlsr = 0 is the result of foreground and background emission within the Galactic disc, while the CMZ itself is responsible for emission over a large range of \vlsr.
The CMZ is a peculiar region of the inner Galaxy that includes a number of large molecular complexes such as \sag\ A ($\ell = 0$, \vlsr\ = 50\,\kms), \sag\ B ($\ell = 0.6\degr$, \vlsr\ = 50\,\kms), \sag\ C ($\ell = 359.3$\degr, \vlsr\ $< 0$\,\kms) and \sag\ D ($\ell = 0.9$, \vlsr\ = 80\,\kms), each covering more than 10 arcmin$^2$; these molecular complexes are labelled in Fig.\,\ref{fig:gc_region}.  This map nicely shows the complex kinematics in the Galactic centre region, in particular the presence of non-circular motions and gas emission at forbidden velocities (negative for $\ell > 0\degr$ and positive for $\ell < 0\degr$; \citealt{riquelme2010} and references therein).

In addition to these two large-scale features, we can also see some finer scale detail such as the narrow absorption  features  at $\ell = 359.5\degr$ with \vlsr $\simeq -50$\,\kms, $-30$\,\kms\ and 0\,\kms; these features are due to absorption of the strong emission emanating from the hot gas in the CMZ by the colder foreground segments of the 3~kpc, Norma, and \sag\ arms (previously observed in HCO$^{+}$ and HCN; e.g. \citealt{fukui1977, fukui1980, linke1981, riquelme2010}). On this plot we have also overlaid the loci of the 3~kpc arms. Comparing the molecular emission with the loci of the near 3~kpc arm (indicated by the lower dashed-dotted yellow line shown in Fig.\,\ref{fig:gc_region}), we find good agreement with the absorption feature seen at $\ell = 359.5\degr$ and $-50$\,\kms, which we have already attributed to this arm.
We also see some association with molecular emission along its length, although this emission is weak and rather sporadic. It is also interesting to note the velocity of the absorption feature associated with the \nor\ arm ($\sim -30$\,\kms), while the model loci of this arm pass very close to \vlsr = 0 at this longitude.

\subsection{Bania molecular clouds}

\begin{figure}
\includegraphics[width=0.49\textwidth]{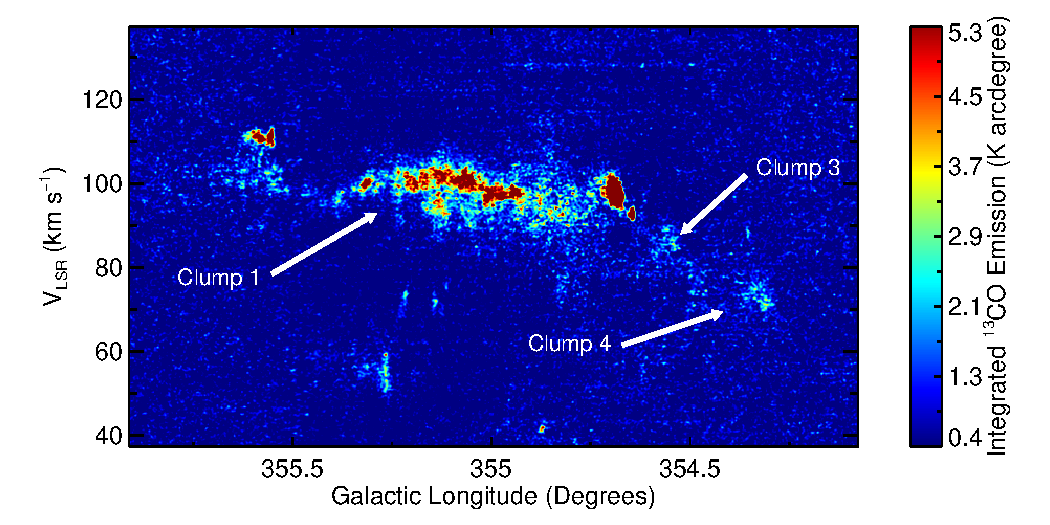}
  \caption{Longitude-velocity map of the Bania Clouds. The features described in the text are labelled following the nomenclature used by \citet{bania1977}. }
    \label{fig:bania_region}
\end{figure}

In Fig.\,\ref{fig:bania_region} we show the Bania complex of molecular clouds \citep{bania1977,Bania+1986}, which consist of three large (40--100~pc) distinct molecular complexes located in a narrow longitude range close to the Galactic centre ($\ell$ between 354.5 and $355.5$\degr) with \vlsr\ velocities of 68, 85 and 100\,\kms.
Adopting the nomenclature from the original papers, these are known as Clump~4, Clump\,3 and Clump\,1, as labelled in Fig.\,\ref{fig:bania_region} (the reference Clump\,2 was given to another object located at $\ell \simeq 3\degr$).
This region is unusual in that it is associated with velocities that are forbidden by Galactic rotation models. Even if this complex was located outside the solar circle on the far-side of the Galaxy at a distance of 40\,kpc, the maximum velocity that we would expect in this direction would be $\sim$16\,\kms\ (\citealt{burton1978}). 

These clouds were originally mapped in $^{12}$CO\,(1-0) where all of the clouds were detected with good signal to noise (\citealt{bania1980, bania1986}). However, in the less abundant \cco\ tracer, Clump\,3 and Clump\,4 are only weakly detected. Clump\,1 is much brighter and appears to be elongated, extending over 1\degr\ in longitude. This cloud is also the only one of the three that is associated with an \hii\ region (G354.67+0.25, \citealt{Caswell1982}), which is located at the western edge of the cloud.

\citet{bania1986} suggested that this complex could be associated with a feature that he refers to as the 135\,\kms\ arm, which can be reproduced by a Galactocentric ring of material with  a radius of 3\,kpc rotating at a velocity of 222\,\kms\ and expanding from the Galactic centre at a speed of 135\,\kms. Clump\,1 is located at the southern terminus of this structure, at a distance of 11.4\,kpc. This large scale structure is not seen in our $\ell v$-map but is clearly seen in the $\ell v$-map of \citet{dame2001} (see Fig.\,2 from \citealt{jones2013}). However, the nature of this 135\,\kms\ arm is contentious (see discussion by  \citealt{jones2013}) and it is not clear if Clump 1 is part of this structure or is entering the dust lane (\citealt{liszt1980}).
Modern simulation efforts often attribute these features to stem from gas approaching the far end of the bar, about to begin the journey back towards the Galactic centre \citep{Baba2010,Li2016,Sormani2018}.

\subsection{Population of nearby wispy clouds}

\begin{figure*}
  \includegraphics[width=0.99\textwidth]{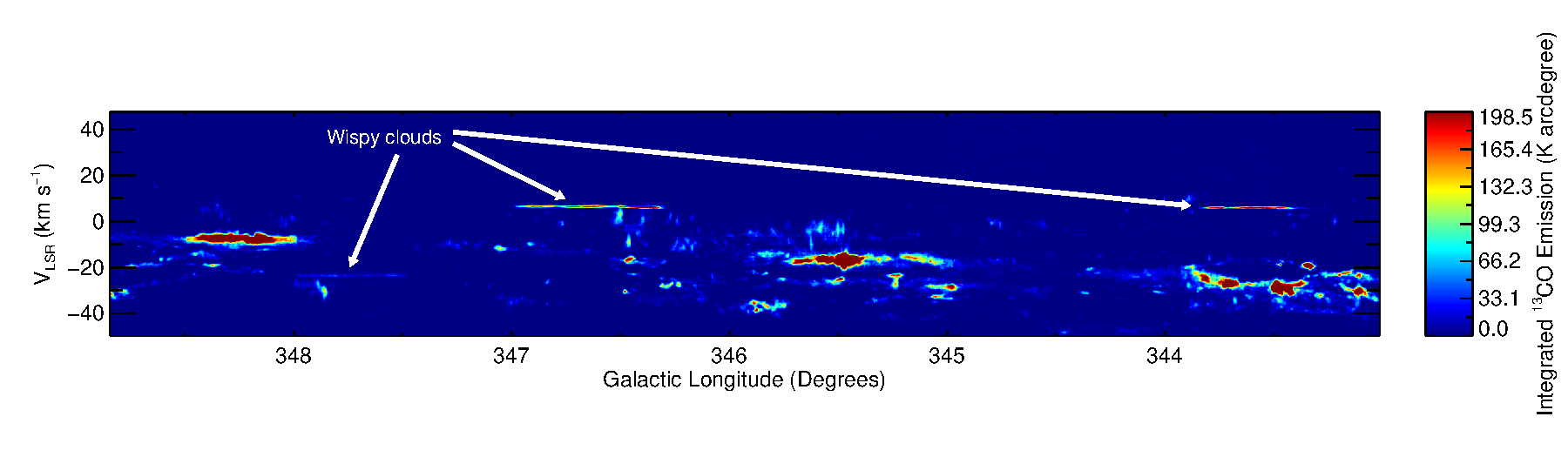}
   \caption{This map is a zoom of a region of the \cco\ $\ell v$-map presented in Fig.\,\ref{fig:lv_plots}, which contains three elongated clouds that have very narrow line-widths (FWHM $\sim$ 0.5-0.75\,\kms). We have classified them as wispy clouds.}
   \label{fig:wispy_lv_map}
\end{figure*}

\begin{figure}
  \includegraphics[width=0.49\textwidth]{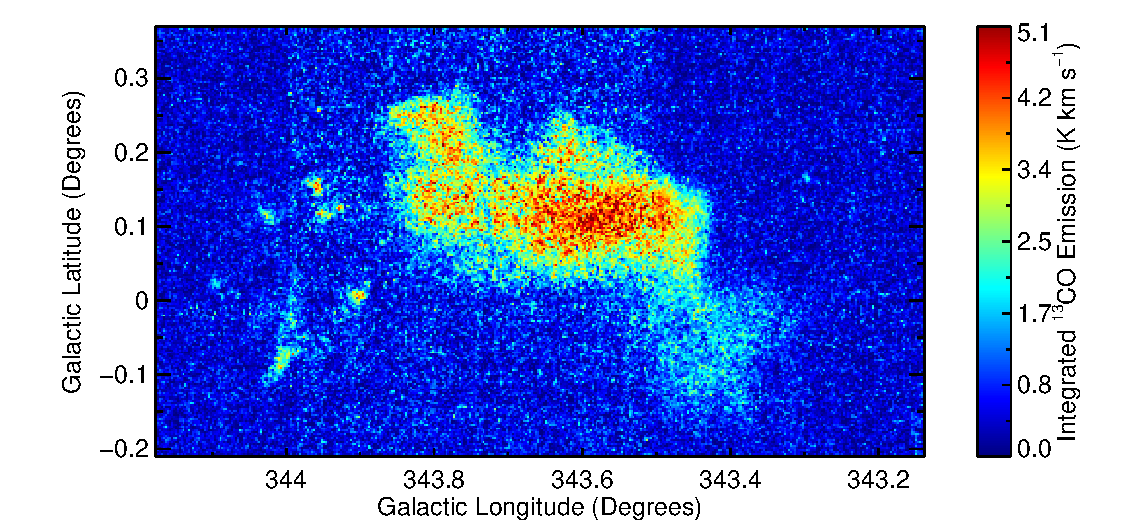}
   \caption{\cco\ emission integrated over the line-width for Cloud\,4 (see Table\,\ref{tab:wispy_clouds} for more details). }
   \label{fig:wispy_lb_map}
\end{figure}

Examination of the $\ell v$-map has revealed the existence of a population of unusual clouds. These appear as very narrow horizontal lines in the $\ell v$-map (see Fig.\,\ref{fig:wispy_lv_map} for some examples), so much so, that we initially thought them to be artificial, perhaps caused by spikes in the spectrometers or due to artefacts introduced during the data reduction procedure. However, on closer examination these were found to be extended over large areas ($\sim$0.5-1\degr\ in diameter) and to have morphologies typical of molecular clouds (see Fig.\,\ref{fig:wispy_lb_map} for an example of their structure).

These clouds have three primary characteristics; they are large in size, they have very narrow line-widths  (FWHM $\sim$0.5-1\,\kms) and they tend to have velocities close to the solar one (i.e. \vlsr\ close to zero). In Table\,\ref{tab:wispy_clouds}, we summarise the positions and velocities for seven of these clouds clearly seen in the $\ell v$-map.
Their velocities and large angular sizes would suggest that the majority of these are local clouds. However, we note that one (Cloud 2) has a velocity that would place it at a larger distance.
Given their narrow line-widths it is possible that these types of clouds have been missed in previous surveys where the velocity resolutions were $>0.5$\,\kms\ as they would only be 1 or 2 velocity channels wide and discarded as artefacts.
It would therefore be interesting to investigate these objects in more detail, however, their near proximity to the Sun makes kinematic distances unreliable and, without these, determining physical properties is not possible.

\begin{figure}
  \includegraphics[width=0.49\textwidth]{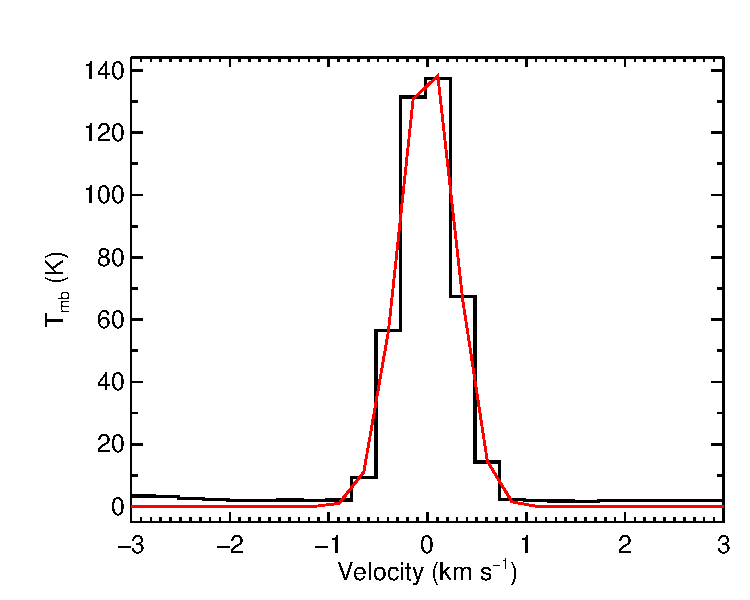}
   \caption{\cco\ emission integrated along the length of  Cloud\,4 located at $\ell = 343.6\degr$ and \vlsr\ = 5.25\,\kms. The black line shows the integrated emission, which has been centred on the \vlsr\ of the source, while the red line shows the result of a Gaussian fit to the profile (FWHM 0.66~\kms, see Table\,\ref{tab:wispy_clouds}).}
   \label{fig:wispy_spectra}
\end{figure}

Typical molecular clouds have FWHM line-widths of a few \kms\ (e.g. Paper\,III) and given that the thermal contribution is of the order of 0.3\,\kms\ (assuming a temperature of 10$-$20\,K) most of the motion in these clouds is non-thermal in nature, and often attributed to turbulence.
The clouds identified in this Section are unusual in that their line-widths are much narrower than typically found for molecular clouds, and, therefore, the thermal and non-thermal components appear roughly balanced.
In Fig.\,\ref{fig:wispy_spectra} we show an example of line-width for Cloud\,4; this has been produced by integrating the emission seen in the $\ell v$-map in longitude (i.e. along its length).
Given that the non-thermal energy can work to support clouds against gravitational collapse, such low values could indicate that these clouds would be potentially unstable to collapse, if they were associated with sufficient mass. In the absence of a robust distance estimate, we cannot determine the masses of the clouds, and thus are limited in our ability to make any further analysis on the nature of these clouds.
Nevertheless, the fact that these are not seen in the C$^{18}$O data suggests that they either have low excitation temperatures or low column densities and are, therefore, rather diffuse and perhaps transient.

\begin{table}
\begin{minipage}{\linewidth}
\begin{center}

\caption{Measured properties of the diffuse clouds.}\label{tab:wispy_clouds}
\begin{tabular}{l....}
\hline

Cloud id.      & \multicolumn{1}{c}{$\ell_{\rm min}$} & \multicolumn{1}{c}{$\ell_{\rm max}$} & \multicolumn{1}{c}{\vlsr} & \multicolumn{1}{c}{FWHM} \\
             & \multicolumn{1}{c}{(\degr)} & \multicolumn{1}{c}{(\degr)} & \multicolumn{1}{c}{(\kms)} &  \multicolumn{1}{c}{(\kms)}\\
\hline
%1 & 3.5 & 4.0 & -64 & 0.53 \\ % 540, 542
1   & 351.0 & 351.6 & 6.0 & 0.65\\ % 821 - 825 %cloud 6681 in catalogue, sigma_v = 0.17
2   & 347.5 & 348.0 & -23.75 & 0.98 \\ %703 - 710 %cloud 6131, sigma_v = 0.37 (quite small... maybe broken into parts?)
3   & 346.3 & 347.0 & 6.75 & 0.75 \\ % 822-829 %cloud 6073 in catalogue, sigma_v = 0.27
4   & 343.4 & 343.8 & 5.25 & 0.66\\ % 822-826 %cloud 5748, sigma_v = 0.32
5   & 342.2 & 342.4 & 5.25 & 0.61 \\ %819 - 822 % couldn't find it. Is this velocity right?
6   & 340.4 & 340.6 & 5.5 & 0.71 \\ % 817 - 823 %cloud 5194 in catalogue, sigma_v = 0.28
7   & 338.6 & 339.2 & -1.5  &  0.96\\ % cloud 5028? v=-2.5 and sigma_v = 0.27
\hline
\end{tabular}

\end{center}
\end{minipage}

\end{table}

We note that the most striking of these clouds are located in the 4$^{\rm{th}}$ quadrant. This potentially highlights a subtle difference between the distribution of molecular gas in the 1$^{\rm{st}}$ and 4$^{\rm{th}}$ quadrants in that there is very little material with a \vlsr\ close to zero in the 1$^{\rm{st}}$ quadrant, and thus fewer local molecular clouds in the portion of the 1$^{\rm{st}}$ quadrant mapped by SEDIGISM than in the 4$^{\rm{th}}$ quadrant. Interestingly, similar clouds with narrow line-widths can also be seen in the $^{13}$CO(3--2) $\ell v$ map of CHIMPS \citep[][Fig.~6]{rigby2016}, for instance a chain of clouds running from +10 to +15~\kms\ in \vlsr\ over the $32.5\degr \leq \ell \leq 35.5\degr$ longitude range.

%%%%%%%%%%%%%%%%%%%%%%%%%%%%%%%%%%%%%%%%%
\section{Other molecular transitions}
\label{sect:exotic_lines}

\begin{figure*}
  \includegraphics[width=0.319\textwidth]{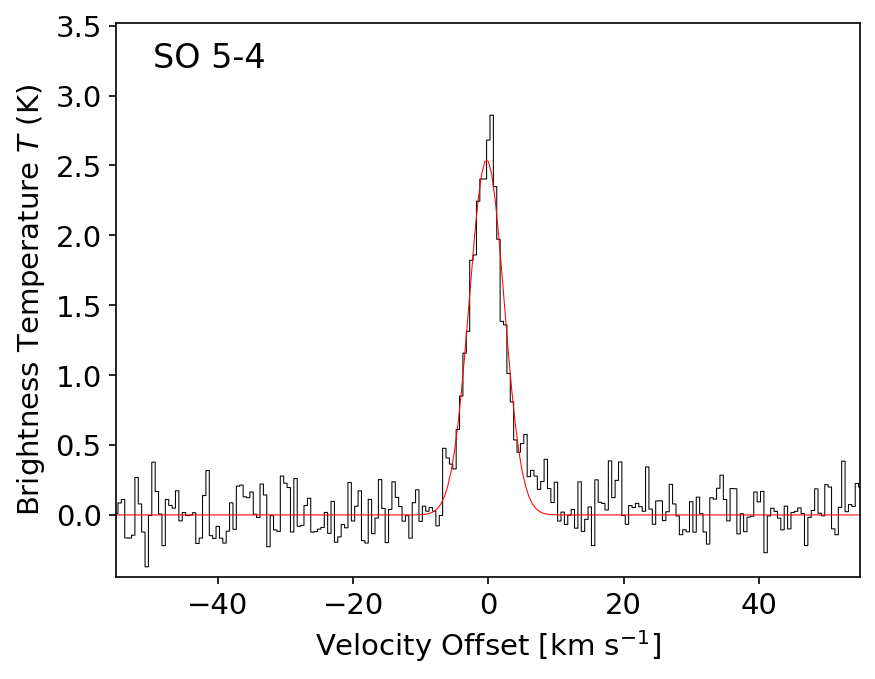}
  \includegraphics[width=0.309\textwidth]{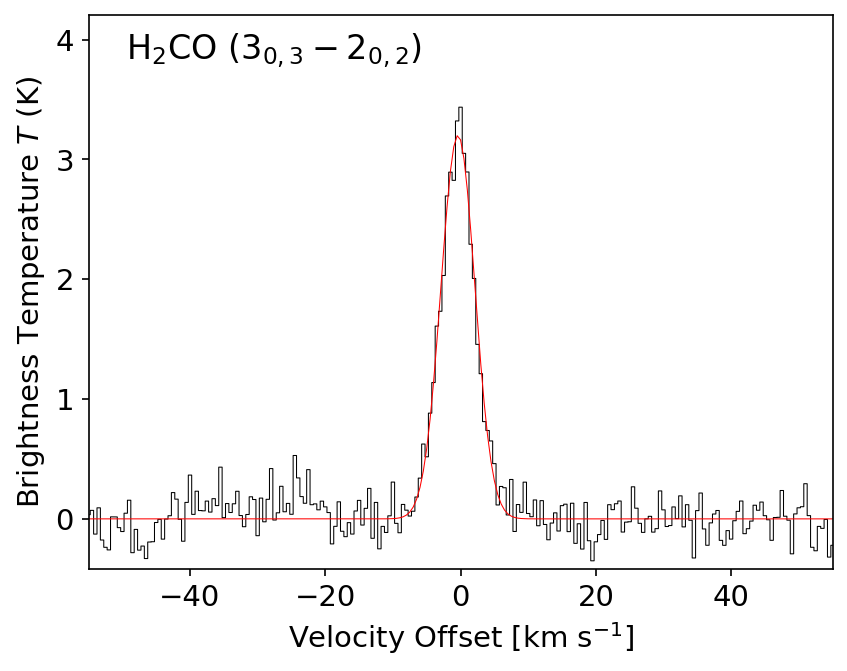}
  \includegraphics[width=0.327\textwidth]{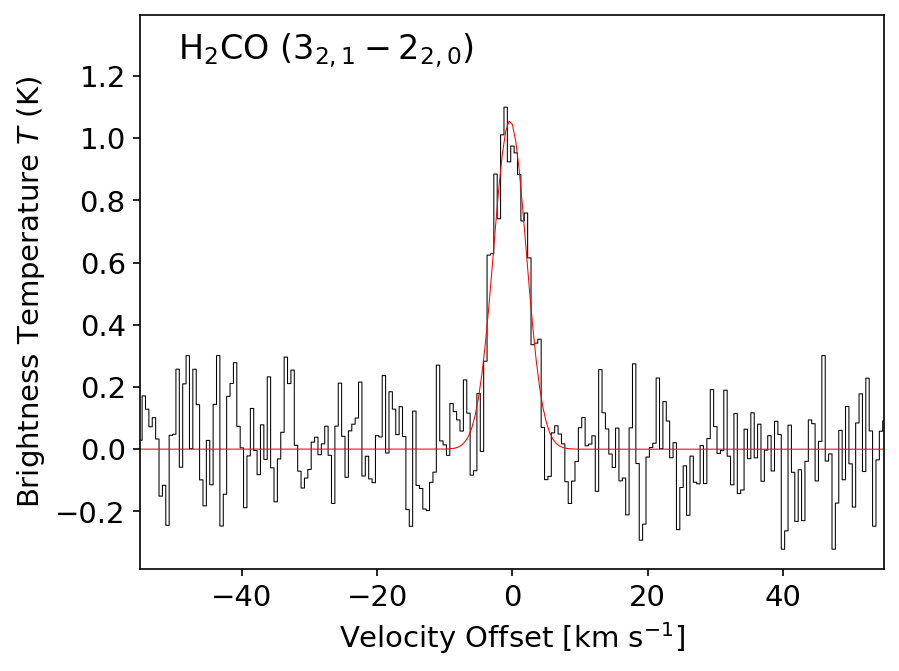}
  \includegraphics[width=0.334\textwidth]{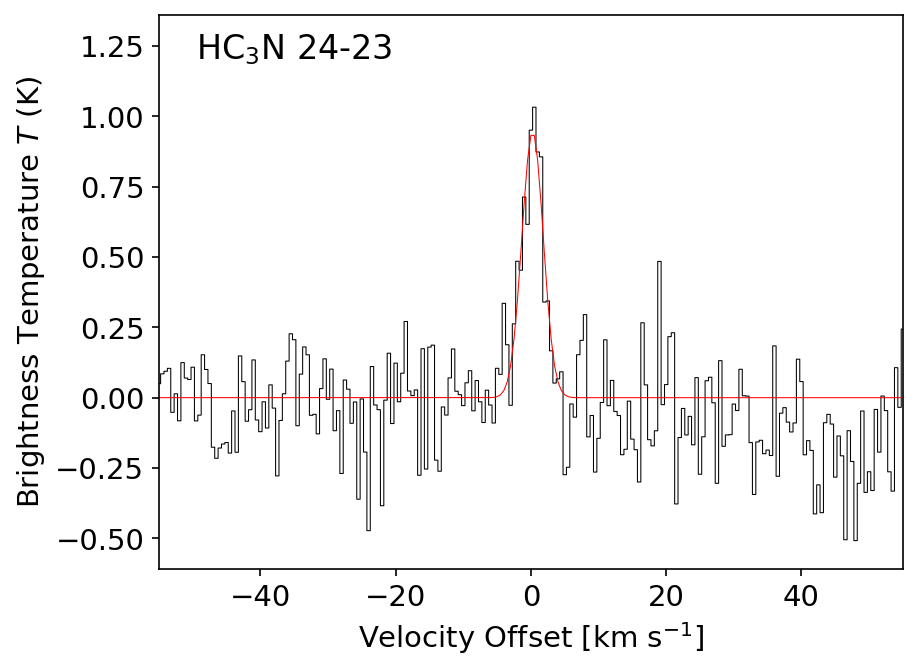}
  \includegraphics[width=0.328\textwidth]{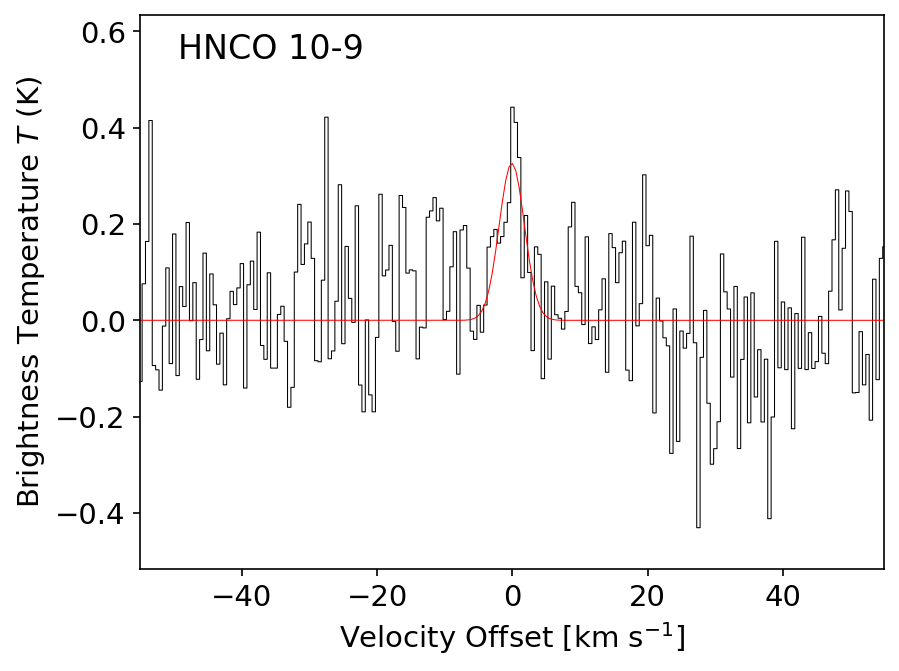}
  \includegraphics[width=0.328\textwidth]{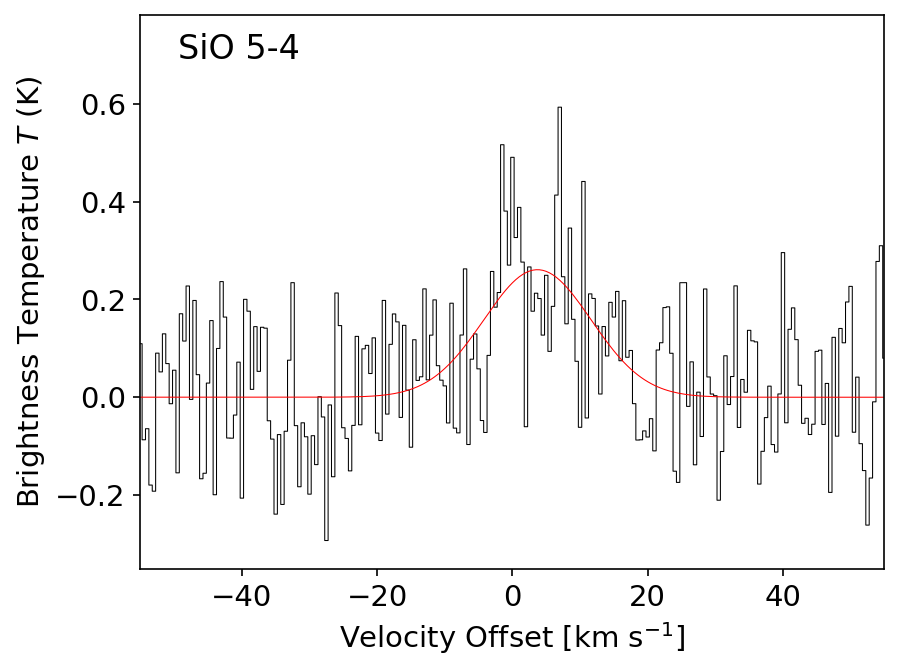}
  \caption{Stacked spectra over the G334 field for the transitions 
   SO(5\,--\,4), H$_2$CO(3$_{0,3}$--2$_{0,2}$), H$_2$CO(3$_{2,1}$--2$_{2,0}$), HC$_3$N(24\,--\,23), HNCO(10\,--\,9) and SiO(5\,--\,4). The red line in each panel shows the best-fit Gaussian profile.}
   \label{fig-stack2}
\end{figure*}

\begin{figure*}
  \includegraphics[width=0.99\textwidth]{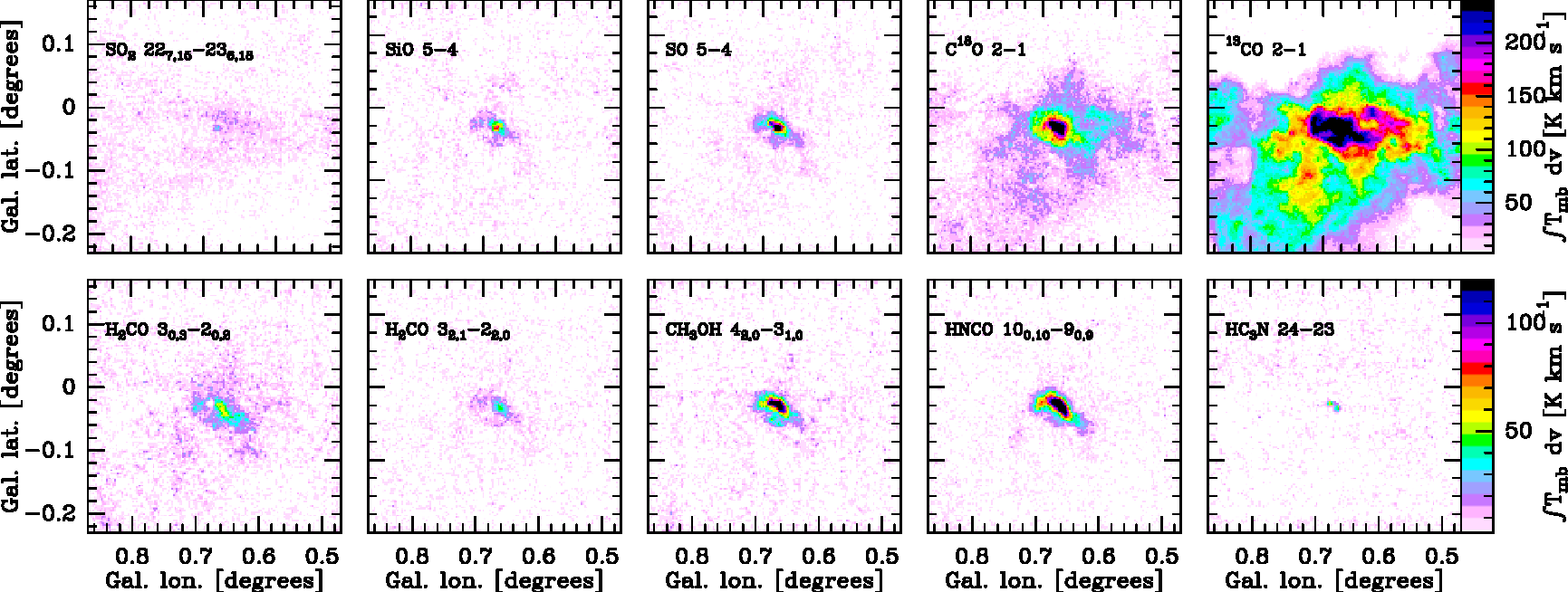}
   \caption{Integrated intensity maps between $+40$ and $+90$\,\kms\   towards a region around Sgr B2 showing extended emission in several molecular tracers. The linear scale corresponds to 3--120 K\,km\,s$^{-1}$ for each panel, except the $^{13}$CO line, where it goes up to 240 K\,km\,s$^{-1}$. The weakest line of H$_2$CO ($3_{2,2}$-$2_{2,1}$ is not shown here.}
   \label{fig-lines}
\end{figure*}

As previously mentioned (Sect.\,\ref{sec-obs}), transitions from several other molecules are included in the spectral tuning used by SEDIGISM, and the data cubes are also publicly available as part of the current data release. This includes spectral cubes for H$_2$CO(3$_{0,3}$-2$_{0,2}$) and ($3_{2,1}-2_{2,0}$), HC$_3$N(24-23), SO($6_5-5_4$), SiO(5\,--\,4) and HNCO(10$_{0,10}$-9$_{0,9}$), with 0.5\,\kms\ velocity resolution. These transitions have lower state energies between 10\,K and 340\,K (cf.\ table\,1 in Paper\,I), and thus probe a wide range of physical conditions. In particular, spectral lines from formaldehyde (H$_2$CO), SO, and cyanoacetylene (HC$_3$N) are commonly detected toward high-mass star-forming regions \citep[e.g.][]{Sutton1985,Belloche2013,Tang2018,Duronea2019}.

These lines are much weaker than the \cco\ and \coo\ lines discussed so far and are likely  only detected towards the densest regions. To explore the utility of these additional lines in our survey we have searched for the possible detection of these molecules in the 2\degr$\times$1\degr\ field centred at 334\degr, which is neither too crowded nor too empty and as such representative of the full survey data. 
Within this field, there are only two positions where non-CO lines are detected: toward the ATLASGAL massive dense clumps AGAL~G333.604$-$00.212 and  AGAL~G333.134$-$00.431 (\citealt{contreras2013}), associated with the the G333 star forming complex.
Towards these two positions, which are also the brightest in \cco\ and \coo\ in this field, we detect H$_2$CO (3$_{0,3}-2_{0,2}$) and ($3_{2,1}-2_{2,0}$), HC$_3$N (24$-$23), and SO ($6_5-5_4$) with peak brightness temperatures of 1$-$3\,K. However, some other, often prominent lines, such as SiO(5\,--\,4) and HNCO (10$_{0,10}-9_{0,9}$), are not detected.

In order to improve our sensitivity to weaker emission we have also performed a stacking analysis to search for emission from other species using {\texttt spectral-cube}\footnote{\url{https://spectral-cube.readthedocs.io/}}.
We analysed the signal-to-noise ratio in the extracted lines vs. intensity threshold in the \cco\ line in steps of 5~K, and we found that most detected lines peaked at a threshold of 30\,K.
Then, we selected all voxels within the \cco\ cube covering the G334 field  where the \cco\ brightness was above a set threshold of 30\,K.
For each pixel in those regions, we used the first moment map to select the peak velocity.
The weak line spectra (i.e., HNCO and SiO) corresponding to these positions were then shifted by the \cco\ velocity and averaged, such that any signal is expected to have velocity offset close to zero. In Fig.\,\ref{fig-stack2} we show the results of our stacking analysis, which reveals that we have indeed detected a weak feature in the HNCO(10\,--\,9) line at ~5-$\sigma$ in integrated intensity.
We have also detected a stronger feature in the SiO transition ($\sim$8-$\sigma$), but this feature is not peaked at 0 \kms\ and is much broader; since high-excitation SiO primarily traces outflows, this broad, offset feature may represent the average of several outflow features over the G334 field.

Using this approach, we can also increase the signal-to-noise of the detection of the more prominent species (e.g. SO, H$_2$CO and HC$_3$N; see Fig.\,\ref{fig-stack2}). The success of this semi-blind stacking approach suggests that detailed studies of Galactic-scale cloud chemistry will be possible despite the survey's relatively short integration times per position.

Finally, we want to highlight that the most extreme star forming regions of our Galaxy exhibit extended emission in several of these weaker lines. We illustrate this in Fig.\,\ref{fig-lines}, where we show an example of the integrated intensity maps of all lines towards a region around Sgr~B2, one of the most active star forming sites in our Galaxy, located close to the Galactic Centre (see Fig.\,\ref{fig:gc_region} for location), where we detect extended emission in the SO (5--4), SiO (5--4), HNCO (10$_{0,10}$-9$_{0,9}$), CH$_3$OH (4$_{2,0}$-3$_{1,0}$) and the H$_2$CO 3$_{0,3}$-2$_{0,2}$ lines.

%%%
\section{Perspectives and conclusions}
\label{sec-perspective}

Here we have presented the first public data release of the SEDIGISM survey, which covers 84~deg$^2$ of the inner Galactic plane in \cco\ and \coo, at 30$''$ angular resolution and a typical noise level of order 1\,K (\tmb) at 0.25~\kms\ resolution.
Future data releases may address remaining issues such as the baseline subtraction in complex regions and other artefacts not fully addressed with the current reduction pipeline (e.g.\ a spike near \vlsr\ -48~\kms\ in \coo\ spectra). 
All data products extracted from this data set, in particular a catalogue and masks of molecular clouds (Paper III) are also being made publicly available alongside this data release, thus providing the community with high added-value products that are complementary to other surveys. This will constitute an invaluable resource for Milky Way studies in the southern hemisphere.

In this paper we have provided an up-to-date description of the data reduction procedure and data products, and highlighted some known issues with some of the fields. We also discussed the Galactic distribution of the molecular gas and investigated its correlation with known star-forming complexes and the large scale structural features of the Galaxy such as the spiral arms. Overall, the data appears consistent with a 4-arm model of the Galaxy.
Using the model from \citet{taylor1993} and updated by \citet{cordes2004}, we found that $\sim$60\,per\,cent of the \cco\ emission is tightly associated with the spiral arms (i.e. within 10\,\kms\ of an arm) and very clear peaks can be seen in the distribution of intensities with Galactocentric distance, that can be attributed to specific spiral arms. 

We have also shown how the velocity information allows us to analyse the complex nature of the molecular gas, which can be separated into different scale structures (filamentary, diffuse and compact structures). This also allows us to investigate the large scale dynamics of the interstellar medium. We have also demonstrated the feasibility of using transitions from less abundant molecular species for science exploitation.
Finally, we have highlighted some interesting regions where the SEDIGISM survey can provide either a new perspective (i.e. Galactic centre region and the Bania molecular complex) or identify a new population of local molecular clouds, which appear as large ($\sim$1\degr) features near \vlsr\ = 0~\kms\ and with very narrow ($<$1~\kms) line-widths.   

A systematic exploitation of the full survey data is now under way. We have also used the SEDIGISM data to confirm the nature of filaments previously identified in ATLASGAL (\citealt{li2016_atlas}) and to explore their kinematics and mass per unit length \citep{mattern2018}.
In parallel to this work we are publishing a catalogue of nearly 11,000 molecular clouds and complexes extracted with SCIMES \citep{ref-scimes}, for which we derived distances and physical properties (Paper\,III), and a systematic assessment of the dense gas fraction and star formation efficiency as a function of environment (Paper\,IV).

In addition to these studies there are a number of ongoing projects aimed at extracting and characterising very long filaments directly from the spectral cubes. We also plan to constrain further the large-scale Galactic structure (number and position of the spiral arms, orientation of the bar), by comparing the SEDIGISM data with the results of simulations. By exploiting the SEDIGISM and ThrUMMS \citep{ref-thrumms} data together, we will also characterise the excitation conditions of the interstellar medium over a large fraction of the Galaxy, extending the exploratory work presented in Paper\,I to the full survey coverage. Other topics under study include the analysis of the dynamical properties and turbulence in giant molecular complexes.

Clearly, a lot of novel studies can be carried out based on the SEDIGISM data alone, and the exploitation of this survey combined with data from other spectroscopic or continuum surveys opens new perspectives for a detailed investigation of the structure and physical conditions of the interstellar medium.

\section*{Acknowledgements}
We thank the anonymous referee for their positive report and constructive comments  .
We are very grateful for the continuous support provided by the APEX staff.
FS acknowledges support from a CEA/Marie Sklodowska-Curie Enhanced Eurotalents fellowship.
ADC acknowledges support from the Royal Society University Research Fellowship URF/R1/191609.
LB acknowledges support from CONICYT project Basal AFB-170002.
HB acknowledges support from the European Research Council under the Horizon 2020 Framework Program via the ERC Consolidator Grant CSF-648505. HB also acknowledges support from the Deutsche Forschungsgemeinschaft (DFG) via Sonderforschungsbereich (SFB) 881 "The Milky Way System" (sub-project B1).
This work was partially funded by the Collaborative Research Council 956 "Conditions and impact
of star formation" (subproject A6), also funded by the DFG.
MW acknowledges funding from the European Union’s Horizon 2020 research and innovation programme under the Marie Sklodowska-Curie grant agreement No 796461. 
TCs has received financial support from the French State in the framework of the IdEx Universit\'e de Bordeaux Investments for the future Program.
CLD acknowledges funding from the European Research Council for the Horizon 2020 ERC consolidator grant project ICYBOB, grant number 818940.
This document was prepared using the Overleaf web application, which can be found at www.overleaf.com.

%%%%%%%%%%%%%%%%%%%%%%%%%%%%%%%%%%%%%%%%%%%%%%%%%%
\section*{Data Availability}

The data presented in this article is available from a dedicated website:
\url{https://sedigism.mpifr-bonn.mpg.de}


\bibliographystyle{mnras}
\bibliography{sedigism_1}

%%%%%%%%%%%%%%%%% AFFILIATIONS %%%%%%%%%%%%%%%%%%%%%%%%%

\bigskip
% List of institutions
{\it \small
\noindent
$^{1}$ Max-Planck-Institut f\"ur Radioastronomie, Auf dem H\"ugel 69, 53121 Bonn, Germany \\
$^{2}$ European Southern Observatory, Alonso de Cordova 3107, Vitacura, Santiago, Chile\\
$^{3}$ Leibniz-Institut f\"ur Astrophysik Potsdam (AIP), An der Sternwarte 16, 14482 Potsdam, Germany \\
$^{4}$ Centre for Astrophysics and Planetary Science, University of Kent, Canterbury CT2 7NH, UK \\
$^{5}$ Laboratoire d'astrophysique de Bordeaux, Univ. Bordeaux, CNRS, B18N, all\'ee Geoffroy Saint-Hilaire, 33615 Pessac, France \\
$^{6}$ School of Physics and Astronomy, Cardiff University, Cardiff CF24 3AA, UK \\
$^{7}$ Department of Astronomy, University of Florida, PO Box 112055, USA \\
$^{8}$ Department of Physics, Faculty of Science, Hokkaido University, Sapporo 060-0810, Japan \\
$^{9}$  West Virginia University, Department of Physics \& Astronomy, P. O. Box 6315, Morgantown, WV 26506, USA \\
$^{10}$ Space Science Institute, 4765 Walnut Street Suite B, Boulder, CO 80301, USA \\
$^{11}$ Observatorio Astrofisico di Arcetri, Largo Enrico Fermi 5, I-50125 Firenze, Italy	\\
$^{12}$ Max-Planck-Institut f\"ur Astronomie, K\"onigstuhl 17, D-69117 Heidelberg, Germany \\
$^{13}$ Departamento de Astronom\'ia, Universidad de Chile, Casilla 36-D, Santiago, Chile \\
$^{14}$ Department of Physics \& Astronomy, University of Exeter, Stocker Road, Exeter, EX4 4QL, United Kingdom \\
$^{15}$ Astrophysics Research Institute, Liverpool John Moores University, 146 Brownlow Hill, Liverpool, L3 5RF, UK \\
$^{16}$ Korea Astronomy and Space Science Institute, 776 Daedeok-daero, 34055 Daejeon, Republic of Korea \\
$^{17}$ Laboratoire d'Astrophysique de Marseille, Aix Marseille Universit\'e, CNRS, UMR 7326, F-13388 Marseille, France \\
$^{18}$ I. Physikalisches Institut, Universit\"at zu K\"oln, Z\"ulpicher Str. 77, D-50937 K\"oln, Germany \\
$^{19}$ Istituto di Astrofisica e Planetologia Spaziali, INAF, via Fosso del Cavaliere 100, I-00133 Roma, Italy\\
$^{20}$ INAF - Istituto di Radioastronomia, Via Gobetti 101, 40129 Bologna, Italy \\
$^{21}$ Astronomy Department, University of Wisconsin, 475 North Charter St, Madison, WI 53706, USA \\
$^{22}$ Dept. of Space, Earth and Environment, Chalmers University of Technology Onsala Space Observatory, 439 92 Onsala, Sweden \\
$^{23}$ Haystack Observatory, Massachusetts Institute of Technology, 99 Millstone Road, Westford, MA 01886, USA \\
$^{24}$ INAF - Osservatorio Astronomico di Cagliari, Via della Scienza 5, 09047 Selargius (CA), Italy \\
$^{25}$ Univ. Grenoble Alpes, CNRS, IPAG, 38000 Grenoble, France \\
$^{26}$ School of Engineering, Macquarie University, NSW 2109, Australia \\
$^{27}$ McMaster University, 1 James St N, Hamilton, ON, L8P 1A2, Canada \\
$^{28}$ European Southern Observatory, Karl-Schwarzschild-Str. 2, D-85748 Garching bei M\"unchen, Germany \\
$^{29}$ Kavli Institute for Astronomy and Astrophysics, Peking University, 5 Yiheyuan Road, Haidian District, Beijing 100871, People's Republic of China}

%%%%%%%%%%%%%%%%%%%%%%%%%%%%%%%%%%%%%%%%%%%%%%%%%%
%%%%%%%%%%%%%%%%% APPENDICES %%%%%%%%%%%%%%%%%%%%%

\appendix

\section{Spectra observed towards reference positions}
\label{app:refspectra}

As described in Sect.~\ref{sec-obs}, position-switching observations were done using a fixed reference position for each map. These reference positions were located at $\pm$1.5\degr\ in galactic latitude. We have done pointed observations in \cco\ (also in position-switching mode) towards all these reference positions, using an off position located one degree further away from the galactic plane (i.e.\ at $\pm$2.5\degr\ in latitude). In some rare cases where we have found that emission was still present in the off position (detectable as an absorption feature in the spectrum), we have repeated the observations using another, nearby off position.

\begin{figure}
	\includegraphics[width=\columnwidth]{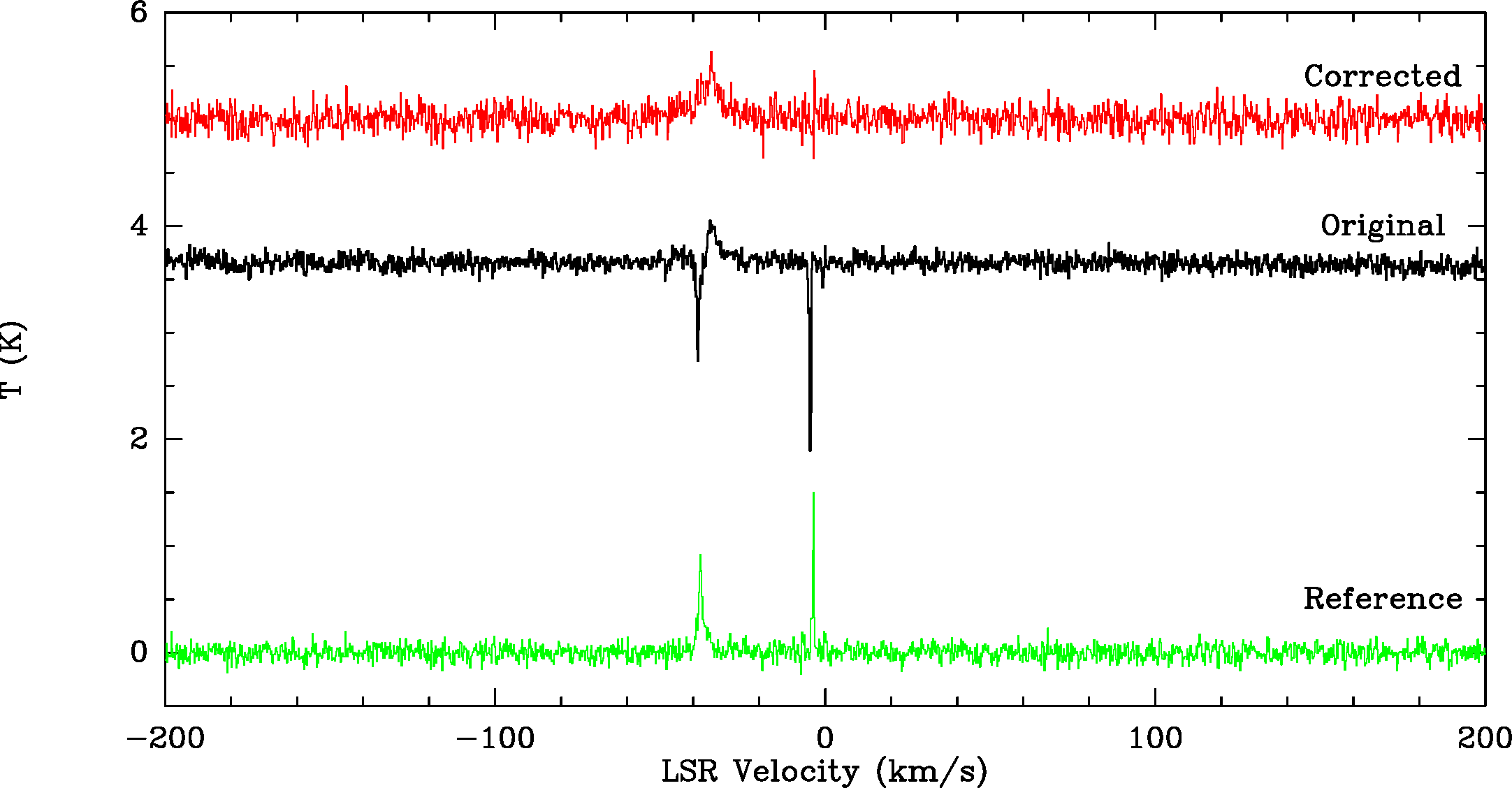}
    \caption{\cco\ spectrum averaged over a small region of the 0.5$\times$0.5~deg$^2$ sub-field 303.25+0.25, before correction (black) and after correction (red).
    The spectrum measured toward the reference position (bottom, shown in green) has been added to each observed spectrum in the data-cube. }
    \label{fig:ex_correction}
\end{figure}

The data were processed using standard procedures in GILDAS/CLASS. The spectra were smoothed to 0.2~\kms\ resolution. We list in Table~\ref{tab:refspectra} the measured rms at that resolution, and the number of lines detected and their \vlsr\ velocities. Only the lines detected at more than 5-$\sigma$ (in integrated intensity) are listed. The spectrum observed towards the reference position was then subtracted from the on-source data only when at least one line was detected.
Fig.~\ref{fig:ex_correction} shows an example case where such a correction was required.

\begin{table*}
\begin{minipage}{\linewidth}
%\begin{center}
\caption{Spectra observed towards the reference positions: measured rms and spectral lines detected. The positions (Col.~1) give the galactic coordinates of the associated on-source maps. The full table is available in the online version of the paper.}\label{tab:refspectra}
\begin{tabular}{lcl}
\hline
Position & $\sigma_{\rm rms}$ [K]  & Lines detected \\
 & ($\delta$V $=$ 0.2~\kms) & \\
\hline
303.25+0.25 	& 0.066 	& two lines: T$_{\rm peak}$ = 0.69 K at \vlsr\ = -37.7~\kms, T$_{\rm peak}$ = 0.85 K at -3.5~\kms 	\\
303.25-0.25 	& 0.066 	& -- 	\\
303.75+0.25 	& 0.066 	& one line: T$_{\rm peak}$ = 0.90 K at
\vlsr\ = -1.1~\kms 	\\
303.75-0.25 	& 0.069 	& -- 	\\
304.25+0.25 	& 0.069 	& three lines: T$_{\rm peak}$ = 0.42 K at \vlsr\ = -40.4~\kms, T$_{\rm peak}$ = 2.60 K at -29.7~\kms, T$_{\rm peak}$ = 0.43 K at -3.3~\kms \\
304.25-0.25 	& 0.068 	& -- 	\\
304.75+0.25 	& 0.069 	& one line: T$_{\rm peak}$ = 1.51 K at \vlsr\ = -26.3~\kms \\
304.75-0.25 	& 0.070 	& -- 	\\
\hline
\end{tabular}
%\end{center}
\end{minipage}
\end{table*}

%%%%%%%%%%%%%%%%%%%%%%%%%
% W43 HERO
%
\section{Comparison between SEDIGISM and HERO data across the W43 region}
\label{A:sed_hero_comp}

In this section we compare the observations of the W43 region from SEDIGISM and from the W43 Hera/EmiR Observations \citep[HERO,][]{Carlhoff2013} project, observed  with the IRAM 30m telescope. Both surveys imaged this region in $^{13}$CO(2-1) and C$^{18}$O(2-1) lines.
The HERO data have better angular (12\arcsec versus 30\arcsec) and spectral (0.15~\kms\ versus 0.25~\kms) resolutions than SEDIGISM, with a similar sensitivity (typically 1~K \tmb\ per 0.15~\kms\ channel, but with smaller pixels; \citealt{Carlhoff2013}).

\begin{figure*}
	\includegraphics[width=0.99\textwidth]{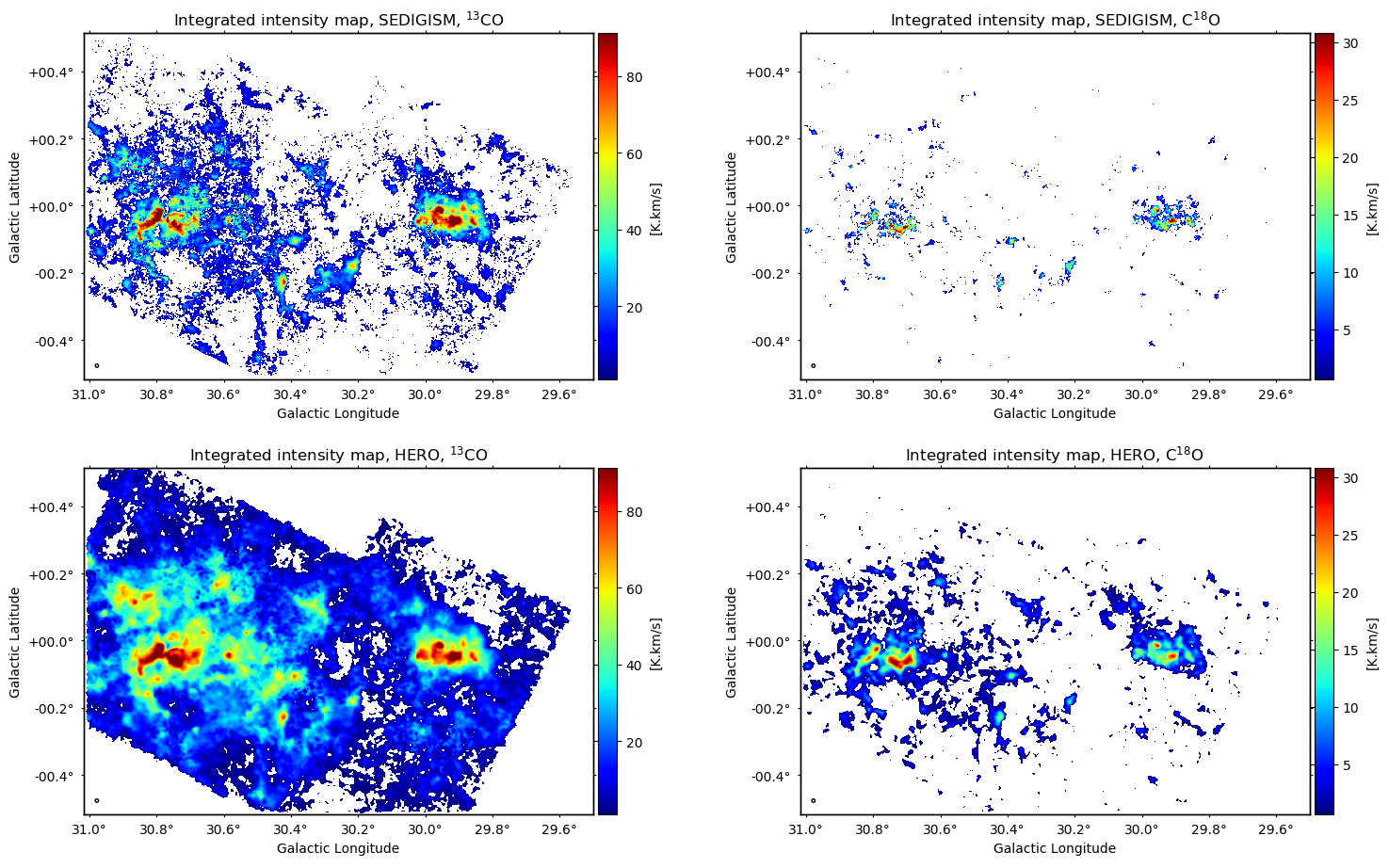}
    \caption{Integrated intensity maps from SEDIGISM (top) and HERO (bottom) $^{13}$CO (left) and C$^{18}$O (right) data. SEDIGISM and HERO data have been homogenized and masked as described in Section~\ref{A:sed_hero_comp}.}
    \label{fig:sed_hero_comp}
\end{figure*}

\begin{figure*}
  \includegraphics[width=0.99\textwidth]{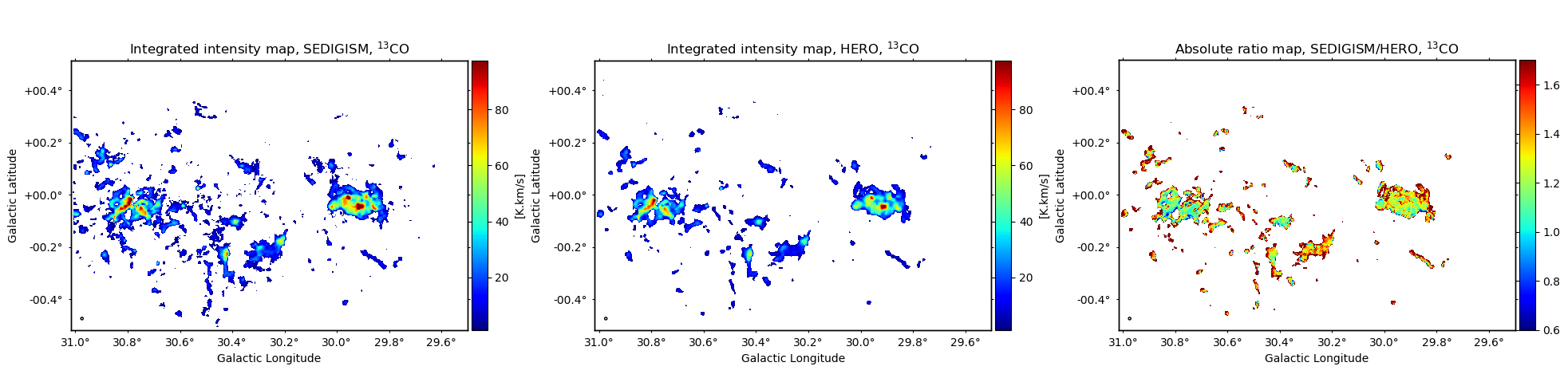}
    \caption{Integrated \cco\ intensity maps from SEDIGISM (left) and HERO (middle), and absolute ratio between the two maps (right). Only voxels above a threshold of 5~K have been considered.}
    \label{fig:sed_hero_13co}
\end{figure*}

To compare the data from the SEDIGISM and HERO surveys we first smoothed and interpolated both data sets to a common resolution of 35" and 0.5~\kms.
To do so we used the python package {\sc spectral\_cube} and the smoothing procedures described in its documentation\footnote{\url{https://spectral-cube.readthedocs.io/en/latest/smoothing.html}}. Since the two data sets do not cover exactly the same region, we regridded the HERO data on the SEDIGISM data grid, using the {\sc astropy} {\sc reproject} function.

To consider only the significant emission in the comparison, we mask the data using a dilate masking technique \citep{roso_leroy}. This technique consists of generating two masks, at relatively low and high signal-to-noise ratio (SNR). Connected regions in the position-position-velocity (PPV) space in the low signal-to-noise ratio mask that do not contain a region of the high signal-to-noise ratio mask are eliminated from the final mask. The result is an actual expansion of the high signal-to-noise regions to lower significant emission, without the inclusion of noisy peaks. For our purposes we consider $SNR\ge 10$ for the high signal-to-noise ratio mask and $SNR\ge 5$ for the low signal-to-noise ratio one. We show the resulting integrated intensity maps (integrated over the full available range of \vlsr) for the two data sets in Fig.~\ref{fig:sed_hero_comp} for $^{13}$CO (left column) and C$^{18}$O (right column).

It is clear from Fig.~\ref{fig:sed_hero_comp}, that the HERO data have a lower noise than SEDIGISM, since less pixels are masked. Beside this, the $^{13}$CO integrated maps from the two surveys appear largely alike (Fig.~\ref{fig:sed_hero_comp}, left column).
To avoid biases due to the different sensitivity in the two surveys, we have also computed integrated $^{13}$CO maps considering only voxels above a fixed value of \tmb \, = \, 5~K (Fig.~\ref{fig:sed_hero_13co}).
The ratio between the two integrated intensity maps appears mostly consistent around unity (within the calibration uncertainty), which demonstrates that both data sets are consistent.
Since the \coo\ line emission is much weaker than $^{13}$CO, we can only compare the brightest regions between both data sets (Fig.~\ref{fig:sed_hero_comp}, right column).
We notice some slight changes for the distribution of the C$^{18}$O emission between SEDIGISM and HERO, but there is no pronounced systematic difference visible.

\label{lastpage}

\end{document}